\documentstyle[12pt]{article}
\title{Matrix integrals, Toda symmetries, Virasoro constraints and orthogonal
polynomials}

\author{M. Adler\thanks{Department of Mathematics,
Brandeis University, Waltham, Mass 02254,
USA. The support of a National Science Foundation grant
\# DMS 95-4-51179 is gratefully acknowledged.}~~~~~P. van
Moerbeke\thanks{Department of Mathematics, Universit\'e de Louvain, 1348
Louvain-la-Neuve, Belgium and Brandeis
University, Waltham, Mass 02254, USA. The support of National 
Science Foundation
\# DMS 95-4-51179, Nato, FNRS and Francqui Foundation grants is
gratefully acknowledged.}}

\date{}
\ifx\Bbb\undefined
  \def\Bbb#1{{\bf #1}}%
\else
  
\fi
  \newcommand{\BZ}{\Bbb Z}
  
  \newcommand{\BC}{\Bbb C}
  \newcommand{\BY}{\Bbb Y}

\newcommand{\BV}{\Bbb V}

\newcommand{\AR}{{\cal A}}
\newcommand{\DR}{{\cal D}}
\newcommand{\LR}{{\cal L}}
\newcommand{\HR}{{\cal H}}
\newcommand{\MR}{{\cal M}}
\newcommand{\rd}{{\cal B}}

\newcommand{\PR}{{\cal P}}
\newcommand{\XR}{{\cal X}}
\newcommand{\pl}{\partial}

\newcommand{\iy}{\infty}
\newcommand{\vp}{\varphi}
\newcommand{\vr}{\varepsilon}
\newcommand{\pp}{\ldots}

\newcommand{\dt}{\delta}
\newcommand{\al}{\alpha}
\newcommand{\be}{\beta}

\newcommand{\Sg}{\Sigma}

\newcommand{\Om}{\Omega}

\newcommand{\lb}{\lambda}

\newcommand{\ga}{\gamma}

\newcommand{\rg}{\rightarrow}
\newcommand{\lrg}{\longrightarrow}

\newcommand{\la}{\langle}
\newcommand{\ra}{\rangle}

\catcode`\@=11
\newskip\centrer \centrer=0pt plus 1000pt minus 1000pt
\def\ALIGNER#1\aligner%
{\let\\=\cr
$$\displ@y \tabskip=\centrer
\halign to\displaywidth{\hfil$\@lign\displaystyle{##}$\tabskip=0pt
&$\@lign\displaystyle{{}##}$\hfil\tabskip=\centrer
&\llap{$\@lign##$}\tabskip=0pt\crcr#1\crcr}$$}
\def\gauche#1{\makebox[0pt][l]{\hskip-\textwidth$#1$}}

\catcode`\@=12

\newcommand{\MAT}[1]{\left(\begin{array}{*#1c}}
\newcommand{\mat}{\end{array}\right)}

\begin{document}
\maketitle

\hspace{2cm} D\'edi\'e avec admiration au Professeur Paul Malliavin
\vspace{1cm}
\newline{\bf Symmetries\footnote{A final version of the present paper has appeared
in: {\em Duke Math. Journal}, {\bf 80}, 863--911 (1995)} of the infinite Toda
lattice.} The symmetries for the infinite Toda lattice, 
$${\frac{\pl  L}{\pl
t_n}=[\frac{1}{2}{( L^n)}_s,L],\,\,\,n=1,2,\pp},\leqno{(0.1)}$$
viewed as isospectral deformations of bi-infinite tridiagonal matrices
$L$, are time-dependent vector fields transversal to the Toda hierarchy;
bracketing a symmetry with a Toda vector field yields another vector field in the
hierarchy. As is well known, the Toda
hierarchy is intimately related to the Lie algebra splitting of
$gl(\infty)$,
$$
gl(\infty)=\DR_s\oplus\DR_b\ni A = A_s + A_b,\leqno{(0.2)}
$$
into the algebras of skew-symmetric $A_s$ and lower
triangular (including the diagonal)  matrices $A_b$ (Borel matrices).
We show that this splitting plays a
prominent role also in the construction of the Toda symmetries and their action on
$\tau-$functions; it also plays a crucial role in obtaining the Virasoro constraints
for matrix integrals and it ties up elegantly with the theory of orthogonal
polynomials .

Define matrices
$\delta$ and $\vr$, with $[\delta,\vr]=1$, acting on characters
$\chi(z)=(\chi_n(z))_{n\in\BZ}=(z^n)_{n\in\BZ}$  as  
$$
\delta\chi=z\chi\mbox{     and    
}\vr\chi=\frac{\pl}{\pl z}\chi~.\leqno{(0.3)}
$$
This enables us to define a wave operator
$S$, a wave vector $\Psi$, 
$$
L=S \delta S^{-1} ~~\mbox{and}~~
\Psi=S\exp^{\frac{1}{2}\sum^\infty_1t_iz^i}\chi(z),\leqno{(0.4)}
$$
and an operator $M$, reminiscent of Orlov and Schulman's $M$-operator for the
KP-equation, such that 
$$
L\Psi=z\Psi\quad\mbox{and}\quad M\Psi=\frac{\pl}{\pl z}\Psi ,\leqno{(0.5)}
$$
thus leading to identities of the form:
$$M^{\be}L^{\al}\Psi=z^{\alpha}(\frac{\pl}{\pl z})^{\be}\Psi.\leqno{(0.6)}
$$

The vector $\Psi$ and the matrices $S,L$ and $M$ evolve in a way, which is
compatible with the algebra splitting above,
$$
\frac{\pl  \Psi}{\pl t_n}=
\frac{1}{2}{(L^n)}_s  \Psi~~ \mbox{and}~~\frac{\pl S}{\pl
t_n}=-\frac{1}{2}{(L^n)}_b S ,\leqno{(0.7)}$$
$${\frac{\pl  L}{\pl
t_n}=[\frac{1}{2}{( L^n)}_s,L]}\quad\mbox{and} \quad {\frac{\pl  M}{\pl
t_n}=[\frac{1}{2}{( L^n)}_s,M]},\leqno{(0.8)}
$$
and the wave vector $\Psi$ has the following representation in terms of a
vector\footnote{with
$\eta=\sum_1^{\iy}\frac{z^{-i}}{i}\frac{\pl}{\pl t_i}$; note the $1/2$ appearing
in $\Psi$  } of
$\tau$-functions $\tau=(\tau_n)_{n\in Z}$:
$$
\Psi(t,z)=e^{\frac{1}{2}\Sg t_iz^i}
\Bigl(z^n\frac{e^{-\eta}\tau_n(t)}{\sqrt{\tau_n(t)\tau_{n+1}(t)}}\Bigr)
_{n\in\BZ}\equiv (z^n\Psi_n)_{n\in\BZ}.\leqno{(0.9)}$$
The wave vector defines a $t$-dependent flag 
$$\dots\supset W^t_{k-1}\supset W^t_k \supset W^t_{k+1}\supset \dots
$$
of nested linear spaces, spanned by functions of z,
$$W^t_k\equiv\mbox{span}\{z^k \Psi_k, z^{k+1} \Psi_{k+1}, \dots\}\leqno{(0.10)}
$$

Formula (0.6) motivates us to give the following definition of symmetry
vector fields (symmetries), acting on the manifold of wave functions $\Psi$
and inducing a Lax pair on the manifold of $L$-operators \footnote{sometimes
$\BY_{M^{\beta}L^{\al}}$ will stand for $\BY_{z^{\al}(\frac{\pl}{\pl z})^{\be}}$}:
$$\BY_{z^{\al}(\frac{\pl}{\pl z})^{\be}}\Psi=-(M^{\be}L^{\al})_b\Psi
\quad \mbox{and}\quad \BY_{z^{\al}(\frac{\pl}{\pl
z})^{\be}}L=[-(M^{\be}L^{\al})_b,L].
\leqno{(0.11)}$$ 
It turns out that {\it only the vector fields}
$$\BY_{\ell,n}:= \BY_{z^{n+\ell}(\frac{\pl}{\pl z})^{n}} =-(M^n
L^{n+\ell})_b, \mbox{for}~ n=0, \ell~
\in \BZ~~\mbox{and}~~ n=1,\ell\geq-1,\leqno{(0.12)}
$$
{\it conserve the tridiagonal nature of the matrices} $L$. The expressions
(0.12), for $n=1$, $\ell<-1$ have no geometrical meaning, as the
corresponding vector fields move you out of the space of tridiagonal matrices.
This phenomenon is totally analogous to the KdV case (or pth Gel'fand-Dickey),
where a certain algebra of symmetries, a representation of the
sub-algebra \footnote {$\mbox{Diff}(S^{1})^+:=\mbox{span} \{ z^{k+1}\frac{\pl}{\pl
z},~k\geq -1
\}$} Diff$(S^1)^+ \subset$ Diff$(S^1)$ of holomorphic
vector fields on the circle, maintains the differential nature of the 2nd order
operator
$\frac{\pl^2}{\pl x^2}+q(x)$ (or pth order differential operator).

According to a
non-commutative Lie algebra splitting theorem, due to ([A-S-V]), stated in section
2 and adapted to the Toda lattice, we have a Lie algebra anti-homomorphism:
$$
w_2\equiv=\{z^{n+\ell}(\frac{\pl}{\pl z})^{n}, n=0,\ell\in\BZ, \mbox{or}~
n=1,\ell\geq-1\}\hspace{6cm}\leqno{(0.13)}$$
$$\hspace{5cm} \rightarrow
\{\mbox{tangent vector fields}~ \mbox{on the}~ \Psi 
\mbox{-manifold}\}:
$$
\medbreak
$$
\hspace{2cm} z^{n+\ell}(\frac{\pl}{\pl z})^{n} \mapsto
\BY_{z^{n+\ell}(\frac{\pl}{\pl z})^{n}},~\quad\mbox{acting on}~ \Psi \mbox{
as in} ~(0.11),
$$ to wit,
$$
[\BY_{\ell,0},\BY_{m,1}]=\ell
\BY_{m+\ell,0}\quad\mbox{and}\quad [{\Bbb Y}_{\ell ,1}, {\Bbb Y}_{m
,1}] = (\ell-m) {\Bbb Y}_{m+\ell ,1}.\leqno{(0.14)}
$$

\bigbreak
{\bf Transferring symmetries from the wave vector to the $\tau$-function
}. An important part of the paper (section 3) is devoted to
understanding how, in the general Toda context, the symmetries $\BY_{\ell,n}$ acting
on the manifold of wave vectors $\Psi$ induce vector fields $\LR_{\ell,n}$ on the
manifold of
$\tau$-vectors $\tau=(\tau_j)_{j\in \BZ}$, for $n=0$ or $1$; this new result
is contained in Theorem 3.2: 

$$
\BY_{\ell,n}\log\Psi
=(e^{-\eta}-1)\LR_{\ell,n}\log\tau + \frac{1}{2}
\Bigl(\LR_{\ell,n}\log(\frac{\tau}{\tau_{\delta}})
\Bigr). \quad\quad({\it Fundamental~ relation})\leqno{(0.15)}
$$
$$
\hspace{6cm}\mbox{for }n=0,~\ell \in
\BZ \mbox{ and }n=1, ~\ell \geq-1
$$

\noindent where `` $\BY_{\ell,n} \log$" and ``$\LR_{\ell,n} \log$" act as
logarithmic derivatives\footnote{for instance 
$
\BY_{\ell,n}\log\Psi_j:=\frac{-\left(( M^n  L^{n+\ell})_b
\Psi\right)_j}{\Psi_j} \quad
 \mbox{and} \quad
\left(\LR_{\ell,n}\log
(\frac{\tau}{\tau_{\delta}})\right)_j\equiv\frac{\LR_{\ell,n}\tau_j}{\tau_j}
-\frac{\LR_{\ell,n}\tau_{j+1}}{\tau_{j+1}}
$}, where 
$
\LR_{\ell,n}f=(\LR^j_{\ell,n}f_j)_{j\in \BZ}$,
and \footnote{set $J^{(0)}_{n}=\delta_{n,0}$, 
$J_n^{(1)} =
\frac{\pl}{\pl t_n}+ \frac{1}{2}(-n)t_{-n}$, and $J_n^{(2)}=\Sg_{i+j=n}
:J_i^{(1)} J_j^{(1)}:$} 
$$
\LR^{j}_{\ell,1} =J^{(2)}_{\ell}
+(2j-\ell-1)J^{(1)}_{\ell}+(j^2-j)J^{(0)}_{\ell},
\,\, ~~\LR^{j}_{\ell,0} = 2 J^{(1)}_{\ell}+2jJ^{(0)}_{\ell}. \leqno{(0.16)}
$$
Note the validity of the relation not
only for infinite matrices, but also for semi-infinite matrices. Also note the
robustness of formula (0.15): it has been shown to be valid in the KP-case
(continuous) and the 2-dimensional Toda lattice (discrete) by [A-S-V].  We give here
an independent proof of this relation, although it could probably have been 
derived from the [A-S-V]-vertex operator identity for the two-dimensional Toda
lattice. The rest of the paper will
be devoted to an application of the fundamental relation. 
\bigbreak

{\bf Orthogonal polynomials, skew-symmetric matrices and Virasoro constraints.}
Consider now in section 4 an orthonormal polynomial basis 
$(p_n(t,z))_{n\geq 0}$ of 
${\cal H}^+\equiv \{1,z,z^2,\dots\}$ with regard to the weight $\rho_0(z)
e^{\sum^\infty_0t_iz^i}dz=e^{-V_0+\sum^\infty_0t_iz^i}$ on the interval
$[a,b],-\infty\leq a<b\leq \infty$, satisfying:
$$ -\frac{\rho'_0}{\rho_0}
=V'_0=\frac{\sum^{\infty}_{0}b_iz^i}{\sum^{\infty}_{0}a_iz^i}=:
\frac{h_0}{f_0}\quad \mbox{and}\quad f_0(a)\rho(a)
a^k=f_0(b)\rho(b) b^k =0~(k=0,1,\ldots)
.\leqno{(0.17)}
$$
The polynomials $p_n(t,z)$ are $t$-deformations of $p_n(0,z)$, through the
exponential in the weight. Then the semi-infinite vector $\Psi$ and semi-infinite
matrices $ L$ and $ M$, defined by
$$
\Psi(t,z)\equiv
\exp^{\frac{1}{2}\sum^\infty_1t_iz^i}\bigl(p_n(t,z)\bigr)_{n\geq 0}
,\quad z\Psi= L\Psi \quad\mbox{and}\quad \frac{\pl}{\pl z}\Psi
=
 M \Psi\leqno{(0.18)}
$$
are solutions of the Toda differential equations (0.7) and (0.8). Moreover,
$ \Psi(t,z)$ can be represented by (0.9) with
$\tau_0=1$ and
$$
\tau_n=\frac{1}{\Omega_n n!}\int_{{\cal
M}_n(a,b)}dZ\,e^{-Tr\,V_0(Z)+\sum_1^{\iy}t_i\,Tr\,Z^i},\quad n\geq
1;\leqno{(0.19)}$$ here the integration is taken over a subspace ${\cal M}_n(a,b)$
of the space of
$n\times n$ Hermitean matrices $Z$, with eigenvalues $\in [a,b].$

We prove in Theorem 4.2 that in terms of the matrices $ L$ and $ M$,
defined in (0.18) and in terms of the anti-commutator
$\{A,B\}:=\frac{1}{2}(AB+BA)$, the semi-infinite matrices\footnote{the matrices
$V_m$ are not to be confused with the potential $V_0$, appearing in the weight. }
$$V_m:=\{ Q, L^{m+1}\}= Q L^{m+1}+\frac{m+1}{2} L^mf_0(
L),~~
\mbox{with}~~V_{-1}= Q:=
 M f_0( L)+\frac{(f_0\rho_0)'}{2\rho_0}( L).\leqno{(0.20)} 
$$
are skew-symmetric for $m
\geq -1$ and form a representation of the Lie algebra of holomorphic
vector fields
$\mbox{Diff}(S^{1})^+$, i.e. they satisfy
$$
[V_m,V_n]=(n-m)\sum_{i\geq 0}a_iV_{m+n+i},\quad m,n\geq -1.
\leqno{(0.21)}
$$
Thus, in terms of the splitting (0.2), we have for orthogonal polynomials 
the following identities: 
$$(V_m)_b =0,\quad \mbox{for
all}\quad m\geq-1,\leqno{(0.22)}$$
leading to the {\it vanishing} of a whole algebra of symmetry vector fields
$\BY_{V_m}$ on the locus of wave functions $\Psi$, defined in (0.18); then using
the fundamental relation (0.15) to transfer the vanishing statement to the
$\tau$-functions
$\tau_n$, we find the Virasoro-type
constraints for the $\tau_n,~n\geq 0$ and for
$m=-1,0,1,2,\dots$: 
$$
\sum_{i\geq
0}\left(
a_i(J_{i+m}^{(2)}+2n\,J_{i+m}^{(1)}+n^2\,J^{(0)}_{i+m})-b_i
(J_{i+m+1}^{(1)}+n\,J^{(0)}_{i+m+1})
\right)\tau_n=0,\leqno{(0.23)}$$
in terms of the coefficients $a_i$ and $b_i$ of $f_0$ and $h_0$ (see
(0.18)). In his fundamental paper [W], Witten had observed, as an incidental 
fact, that in the case of Hermite polynomials $V_{-1}= M- L$ is a skew-symmetric
matrix. In this paper we show that skew-symmetry plays a crucial role; in fact the 
{\it Virasoro constraints} (0.23) are tantamount to {\it the skew-symmetry of the
semi-infinite matrices} (0.20). 

They can also be obtained, with uninspired tears, upon substituting $$
Z\mapsto Z+\vr f_0(Z) Z^{k+1} \leqno{(0.24)}
$$
in the integrand of (0.19); then the linear terms in $\vr$ in the integral (0.19)
must vanish and yield the same Virasoro-type constraints (0.23) for each of the
integrals
$\tau_n$ as is carried out in the appendix.

At the same time, the methods above solve the ``{\it string
equation}", which is : for given $f_0$, find two semi-infinite matrices, a
\underline {symmetric}
$ L$ and a \underline {skew-symmetric} $ Q$, satisfying
$$[ L, Q]=f_0( L). \leqno{(0.25)}$$

For the {\bf classical orthogonal polynomials}, as explained in section 6, the
matrices
$ L$ and
$ Q$, matrix realizations of the operators $z$ and 
$\sqrt{\frac{f_0}{\rho _0}}\frac{\pl}{\pl z}\sqrt{\rho_0
f_0}$ respectively, acting on the space of polynomials, are both tridiagonal, with
$ L$ symmetric and $ Q$ skew-symmetric. In addition, the matrices $ L$
and
$ Q$ stabilize the flag, defined in (0.10), in the following sense:
$$
zW_k \subset W_{k-1}\quad \mbox{and} \quad T_{-1}W_k \subset
W_{k-1}.\leqno{(0.28)}
$$This result is related to a classical Theorem of Bochner; see [C].

The results in this paper
have been lectured on at CIMPA (1991), Como, Utrecht (1992) and Cortona (1993); see 
the lecture notes [vM]. Grinevich, Orlov and Schulman made a laconic remark in a
1993 paper [GOS, p. 298] about defining symmetries for the Toda
lattice. We thank A. Gr\"unbaum, L. Haine, V. Kac, A.
Magnus, A. Morozov, T. Shiota, Cr. Tracy and E. Witten for  conversations, regarding
various aspects of this work. We also thank S. D'Addato-Mu\"es for unscrambling an
often messy manuscript.

\vspace{1cm}

\noindent Table of contents:

\bigbreak

\noindent 1. The Toda lattice revisited

\noindent 2. Symmetries of the Toda lattice and the $w_2$-algebra

\noindent 3. The action of the symmetries on the $\tau$-function

\noindent 4. Orthogonal polynomials, matrix
integrals, skew-symmetric matrices and Virasoro
constraints

\noindent 5. Classical orthogonal polynomials

\noindent 6. Appendix: Virasoro constraints via the integrals

\section{The Toda lattice revisited}

On $\BZ$ we define the function $\chi$ (character)
$$
\chi :\BZ\times\BC\rg\BC :(n,z)\mapsto\chi_n(z)=z^n
$$
and the matrix-operators $\delta$ and $\vr$
$$\delta =
\left(\begin{tabular}{llll}
$\ddots$& & & \\
 &\mbox{\begin{tabular}[b]{cc|cc}
0&1& &\\
 &0&1&\\
\hline
 & &0&1\\
 & & &0
\end{tabular}}&1& \\
 & &0& \\
 & & &$\ddots$ 
\end{tabular}
\right)\mbox{  and  }\vr=\left(\begin{tabular}{llll}
$\ddots$& & & \\
-2&\mbox{\begin{tabular}[t]{cc|cc}
0& & &\\
-1&0& &\\
\hline
 &0&0& \\
 & &1&0
\end{tabular}}& & \\
 &\hfill 2&0&\\
 & &   &$\ddots$ 
\end{tabular}
\right)\leqno{(1.1)}
$$
acting on $\chi$ as
$$
\delta\chi=z\chi\mbox{     and    
}\vr\chi=\frac{\pl}{\pl z}\chi;\leqno{(1.2)}
$$
they satisfy
$$
[\delta,\vr]=1.
$$

\proclaim Proposition 1.1. The infinite
matrix\footnote{$\DR_{k,\ell}$ ($k<\ell\in\BZ$) denotes the set of band
matrices with zeros outside the strip $(k,\ell).$ The symbols
$(\,)_+,(\,)_-,(\,)_0$ denote the projection of a matrix onto $\DR_{0,\iy}$,
$\DR_{-\iy,-1}$ and $\DR_{00}$ respectively.}
$$
\tilde L=\sum_{j\leq 1}a_j\delta^j\in\DR_{-\iy,1},\hspace{1cm}
a_j(t)=\mbox{ diag}(a_j(n,t))_{n\in\BZ}, a_1=1,\leqno{(1.3)}
$$
subjected to the deformation equations
$$
\frac{\pl\tilde L}{\pl t_n}=[(\tilde L^n)_+,\tilde 
L]=[-(\tilde L^n)_-,\tilde L],\hspace{1cm}n=1,2,\pp\,\,.\leqno{(1.4)} $$
has, for generic initial conditions, a representation in terms of  $\tau$-functions
$\tau_n$ 
$$
\tilde L=\tilde S\delta\tilde S^{-1}=\Bigl(\pp,(\frac{\ga_n}
{\ga_{n-1}})^2,\frac{\pl}{\pl t_1}\log
\ga^2_n,1,0,\pp\Bigr)_{n\in\BZ},~ \leqno{(1.5)}
$$
where\footnote{The $p_k$'s are the elementary Schur polynomials
$e^{\sum_1^{\iy}t_iz^i}=\sum_0^{\iy}p_k(t)z^k$ and $p_k(-\tilde\pl):=
p_k(-\frac{\pl}{\pl t_1},-\frac{1}{2}\frac{\pl}{\pl t_2},-
\frac{1}{3}\frac{\pl}{\pl t_3},\pp).$ Also
$[\al]:=(\al,\frac{\al^2}{2},\frac{\al^3}{3},...).$} 
$$\ga_n=\sqrt{\frac{\tau_{n+1}}{\tau_n}},\quad \tilde S=\frac{\tau(t-[\delta^{-1}]
)}{\tau(t)}=\frac{\sum_{n=0}^{\iy}p_n(-\tilde\pl)\tau(t)\delta^{-n}}{\tau(t)}
\leqno{(1.6)}$$
The wave operator $\tilde S$ and the wave vector\footnote{set
$\eta=\sum_1^{\iy}\frac{z^{-i}}{i}\frac{\pl}{\pl t_i}$ and
$\Sg =\sum_1^{\iy}t_jz^j$}  
$$
\tilde\Psi:=e^{\sum_1^{\iy}t_iz^i}\tilde
S\chi(z)=\Bigl(z^ne^{\Sg}\frac{e^{-\eta}\tau_n}{\tau_n}\Bigr)_{n\in\BZ}
=:(z^n\tilde\Psi_n)_{n\in\BZ}
\leqno{(1.7)} $$
satisfy
$$\tilde L \tilde \Psi=z \tilde \Psi,~~\frac{\pl\tilde \Psi}{\pl t_n}=(\tilde
L^n)_+\tilde
\Psi,~~
\frac{\pl\tilde S}{\pl t_n}=-(\tilde L^n)_-S.\leqno{(1.8)} 
$$

\medbreak

\noindent{\it Proof}: The proof of this statement can be deduced from the
work of Ueno-Takasaki [U-T]; 
we consider only a few points: from equation (1.6), it follows that:
$$
\tilde S= I-A\dt^{-1}-B\dt^{-2}-\pp,\quad\mbox{and}\quad\tilde
S^{-1}=I+A\dt^{-1}+B\dt^{-2}+A\dt^{-1}A\dt^{-1}+\pp\,
.\leqno{(1.9)}$$
with
$$\quad A=\frac{\pl}{\pl t_1}\log\tau,~~\mbox{and}~~
B=-\frac{p_2(-\tilde\pl)\tau(t)}{\tau(t)}.
$$
Then, calling $p_k\tau:= p_k(-\tilde\pl)\tau$, we have
\medbreak
\noindent (1.10)
\begin{eqnarray*}
\tilde L=\tilde S\dt\tilde
S^{-1}&=&\dt+(A_{n+1}-A_n)_{n\in\BZ}\dt^0+(B_{n+1}-B_n+
A_nA_{n+1}-A^2_n)_{n\in\BZ}\dt^{-1}+\pp\\
&=&\dt+\Bigl(\frac{\pl}{\pl
t_1}\log\frac{\tau_{n+1}}{\tau_n}\Bigr)_{n\in\BZ}\dt^0\\
& &\,\,\,+\Bigl(
-\frac{p_2\tau_{n+1}}{\tau_{n+1}}+\frac{p_2\tau_n}{\tau_n}+\frac{p_1\tau_n}
{\tau_n}\frac{p_1\tau_{n+1}}{\tau_{n+1}}-(\frac{p_1\tau_n}{\tau_n})^2
\Bigr)_{n\in\BZ}\dt^{-1}+\pp\, ,
\end{eqnarray*}
yielding the representation (1.5), except for the $\delta^{-1}$-term, which we
discuss next. 

To the Toda problem is associated a flag of nested planes
$\tilde W_{n+1}\subset \tilde W_n\in Gr_n$,
\begin{eqnarray*}
\tilde W_n&\equiv&\mbox{
span}\{z^n\tilde\Psi_n,z^{n+1}\tilde\Psi_{n+1},\pp\}\\
&=&\mbox{
span
}z^n\{\tilde\Psi_n,\frac{\pl}{\pl
t_1}\tilde\Psi_n,(\frac{\pl}{\pl t_1})^2\tilde\Psi_n,\pp\}.
\end{eqnarray*}
The inclusion $\tilde W_{n+1}\subset \tilde W_n$ implies, by noting
$\tilde\Psi_k=1+O(z^{-1})$, that $$
z\tilde\Psi_{n+1}=\frac{\pl}{\pl
t_1}\tilde\Psi_n-\al
\tilde\Psi_n\quad\mbox{for some}\quad\al=\al(t).\leqno{(1.11)} $$
Then $\al(t)=\frac{\pl}{\pl t_1}\log\tau_{n+1}/\tau_n$ and putting this
expression in (1.11) yields\footnote{$\{f,g\}=\frac{\pl f}{\pl t_1}g-f
\frac{\pl g}{\pl t_1}$}
$$
\{\tau_n(t-[z^{-1}]),\tau_{n+1}(t)\}+z\Bigl(\tau_n(t-[z^{-1}])\tau_{n+1}(t)-
\tau_{n+1}(t-[z^{-1}])\tau_n(t)\Bigr)=0.
$$
Expanding this expression in powers of $z^{-1}$ and dividing the coefficient
of $z^{-1}$ by $\tau_n\tau_{n+1}$ yield
$$
-\frac{p_2\tau_{n+1}}{\tau_{n+1}}+\frac{p_2\tau_n}{\tau_n}+
\frac{p_1\tau_n}{\tau_n}\frac{p_1\tau_{n+1}}{\tau_{n+1}}-
\frac{\frac{\pl^2\tau_n}{\pl t_1^2}}{\tau_n}=0.
$$
Combining this relation with the customary Hirota bilinear relations,
the simplest one being:
$$
-\frac{1}{2}\frac{\pl^2}{\pl
t_1^2}\tau_n\circ\tau_n+\tau_{n-1}\tau_{n+1}=0,\quad\mbox{i.e.}\quad
\frac{\pl^2}{\pl
t_1^2}\log\tau_n=\frac{\tau_{n-1}\tau_{n+1}}{\tau^2_n},
 $$
we find
$$
-\frac{p_2\tau_{n+1}}{\tau_{n+1}}+\frac{p_2\tau_n}{\tau_n}+
\frac{p_1\tau_n}{\tau_n}\frac{p_1\tau_{n+1}}{\tau_{n+1}}-
(\frac{p_1\tau_n}{\tau_n})^2=\frac{\pl^2}{\pl t_1^2} \log
\tau_n=\frac{\tau_{n-1}\tau_{n+1}}{\tau^2_n},
$$
and thus the representation (1.5) of $\tilde L$,
ending the proof of Proposition 1.1.

\vspace{1cm}

Henceforth, we assume $\tilde L$ as in proposition 1.1, but in addition {\it
tridiagonal}:
$\tilde L=\sum_{-1\leq j\leq 1}a_j\dt^j;$ this submanifold is invariant under the
vector field (1.4); indeed, more generally  if $\tilde L=\sum_{j\leq
1}a_j\dt^j$ is a $N+1$ band matrix, i.e. $a_j=0$ for $j\leq -N\leq 0$, then
$\tilde L$ remains a $N+1$ band matrix under the Toda vector
fields. Moreover consider the Lie algebra
decomposition, alluded to in (0.2), of $gl(\infty)=\DR_s\oplus\DR_b\ni A = (A)_s +
(A)_b$ in skew-symmetric plus lower Borel part (lower triangular, including the
diagonal).

\proclaim Theorem 1.2. Considering the submanifold of tridiagonal matrices $\tilde
L$ of proposition 1.1 and remembering the form of the diagonal matrix
$\ga=(\ga_n)_{n\in\BZ},$ with $\ga_n=\sqrt{\frac{\tau_{n+1}}{\tau_n}},$ we define
a new wave operator $S$ and wave vector $\Psi$ :
$$
S:=\gamma^{-1}\tilde S ~~\mbox{and}~~\Psi:=S\chi(z)e^{\Sigma/2};\leqno{(1.12)}
$$
also define
$$L:=S\delta S^{-1}~~\mbox{and}~~M:= S\bigl(  \vr + \frac{1}{2} \sum^{\iy}_1 k
t_k \delta^{k-1}\bigr)S^{-1}.\leqno{(1.13)}$$
Then the tridiagonal matrix
$$
L=\Bigl(\pp,0,\frac{\ga_n}{\ga_{n-1}},\frac{\pl}{\pl t_1}\log
\ga^2_n,\frac{\ga_{n+1}}{\ga_n},0,\pp\Bigr)_{n\in\BZ}\leqno{(1.14)}
$$
is symmetric and 
$$
\Psi=\ga^{-1}\tilde\Psi=e^{\Sg/2}\ga^{-1}\chi\frac{e^{-\eta}\tau}{\tau}=:(z^n\Psi_n)_{n\in\BZ}\mbox{
with
}\Psi_n:=e^{\Sg/2}\frac{e^{-\eta}\tau_n}{\sqrt{\tau_n\tau_{n+1}}}.\leqno{(1.15)}
$$
The new quantities satisfy:
$$ \frac{\pl \log \ga}{\pl t_n}=\frac{1}{2}{(L^n)}_0~, \quad\frac{\pl S}{\pl
t_n}=-\frac{1}{2}{(L^n)}_b S\quad\mbox{and}\quad
\frac{\pl  \Psi}{\pl t_n}=\frac{1}{2}{(L^n)}_s  \Psi,\leqno{(1.16)}
$$
$$L\Psi=z\Psi~~\mbox{and}~~M\Psi=\frac{\pl}{\pl z}\Psi,\quad \mbox{with}\quad
[L,M]=1,\leqno{(1.17)}$$ and
$$
\frac{\pl L}{\pl t_n}=[\frac{1}{2}(L^n)_s,L],\quad\frac{\pl M}{\pl t_n}=
[\frac{1}{2}(L^n)_s,M].\leqno{(1.18)}
$$

\medbreak

{\sl Proof of Theorem 1.2:} For a given initial condition $\gamma'(0)$,
the system of partial differential equations in
$\gamma'$
$$
\frac{\pl}{\pl t_k}\log\ga '
=\frac{1}{2}(\tilde L^k)_0;\leqno{(1.19)} 
$$
has, by Frobenius theorem, a unique solution, since

\ALIGNER
\left(\frac{\pl}{\pl t_k}
\tilde L^n-\frac{\pl}{\pl
t_n}\tilde L^k\right)_0&=-[(\tilde L^k)_-,\tilde L^n]_0+[\tilde
L^k,(\tilde L^n)_+]_0&\gauche{(1.20)}\\ &=-[(\tilde L^k)_-,(\tilde
L^n)_+]_0+[(\tilde L^k)_-,(\tilde L^n)_+]_0=0,\\
\aligner 
using $[A_+,B_+]_0=0$ and $[A_-,B_-]_0=0$ for arbitrary
matrices $A$ and $B.$ 
Given this solution $\ga'(t)$, define $S':=\ga '^{-1} \tilde S$ and $L':=S' \delta
S'^{-1}=\ga'^{-1} \tilde L \ga'$, and using $\pl \tilde S/\pl t_n=-(\tilde
L^n)_-\tilde S$, compute
 \ALIGNER
\frac{\pl  S'}{\pl t_k} = \frac{\pl(\ga'^{-1} \tilde S)}{\pl
t_k} & =  \frac{\pl \ga'^{-1}}{\pl t_k} \tilde S +
\ga'^{-1}\frac{\pl \tilde S}{\pl t_k} \\
& =  -\ga'^{-1}(\frac{\pl \ga'}{\pl t_k}) \ga'^{-1} \tilde S +
\ga'^{-1}\frac{\pl \tilde S}{\pl t_k}  \\ 
& =  -\ga'^{-1}\frac{1}{2} (\tilde L^k)_0 \tilde S -
\ga'^{-1} (\tilde L^k)_- \tilde S \\
& =  - \frac{1}{2} ( L'^k)_0  S' - 
( L'^k)_-  S' \quad \mbox{since}~ L'=\ga'^{-1} \tilde L \ga'~\mbox{and}~S':=\ga
'^{-1} \tilde S \\ & =  -\left(( L'^k)_- + \frac{1}{2}( 
L'^k)_0\right)  S' = -\frac{1}{2} ( L'^k)_{b'}
 S',
\aligner
where 
$$ A_{b'}:= 2A_- +A_0 \quad \mbox{and}\quad
A_{s'}:=A-A_{b'}.$$ 
It follows at once that
$$\frac{\pl L'}{\pl
t_n}=[-\frac{1}{2}(L'^n)_{b'},L']=[\frac{1}{2}(L'^n)_{s'},L']\leqno{(1.21)} $$

Observe now that the manifold of symmetric tridiagonal matrices $A$ is invariant
under the vector fields
$$\frac{\pl A}{\pl
t_n}=[-\frac{1}{2}(A^n)_{b'},A]=[\frac{1}{2}(A^n)_{s'},A],\leqno{(1.22)} $$
since for $A$ symmetric, the operations $()_{b'}$ and $()_{s'}$ coincide with the
decomposition (0.2):
$$A_{b'}=A_b ~~\mbox{and}~~ A_{s'}=A_s.$$

Now according to formula (1.5), picking  
$\ga'(0):=\ga(0)=\sqrt{\frac{\tau_{n+1}(0)}{\tau_n(0)}}$ as initial condition
for the system of pde's (1.19), makes $L'(0)=\ga'(0)^{-1} \tilde L(0) \ga(0)'$
symmetric. Since $L'$ was shown to evolve according to (1.21) or (1.22), and since
its initial condition is symmetric, the matrix
$L'$ remains symmetric in $t$. Since $L'(t):=\ga'^{-1}(t) \tilde L
(t)\ga'(t)$ is symmetric and since, by definition, $L(t):=\ga^{-1}(t) \tilde L
(t)\ga(t)$ is also symmetric, we have
$L'(t)=L(t)$, and thus
$\gamma'(t)=c\gamma (t)$ for some constant $c$; but $c$ must be $1$, since
$\gamma'(0)=\gamma (0)$. This proves (1.14), (1.16) and the first halfs of (1.17)
and (1.18).

Besides multiplication of $\Psi$ by $z$, which is represented by the matrix $L$,
we also consider differentiation $\pl/\pl z$ of
$\Psi$, which we represent by a matrix $M$:
\ALIGNER
\frac{\pl   \Psi}{\pl z} 
&=e^{\Sg/2}   S \frac{\pl}{\pl z} \chi +
\frac{1}{2}(\sum^{\iy}_1 k t_k z^{k-1}) e^{\Sg/2} S
\chi&\gauche{(1.23)}\\
&=: \left(  P + \frac{1}{2} \sum^{\iy}_1 k t_k
L^{k-1} \right)   \Psi=:   M \Psi
\aligner
with $$[L,M]=1~~\mbox{and}~~P:= S\vr S^{-1}\in{\cal D}_{-\iy,-1}.\gauche{(1.30)}$$
Finally compute 
\begin{eqnarray*}
\frac{\pl M}{\pl t_n}&=&\frac{\pl S}{\pl t_n}(S^{-1}S)(\vr +\frac{1}{2}\Sg
kt_k\dt^{k-1})S^{-1}\\
& &+\frac{1}{2}nL^{n-1}-S(\vr+\frac{1}{2}\Sg kt_k\dt^{k-1})S^{-1} \frac{\pl
S}{\pl t_n} S^{-1}\\
&=&-\frac{1}{2}(L^n)_bM+[\frac{1}{2}L^n,M]+\frac{1}{2}M(L^n)_b\\
&=&[-\frac{1}{2}(L^n)_b+\frac{1}{2}L^n,M]\\
&=&[\frac{1}{2}(L^n)_s,M],
\end{eqnarray*}
ending the proof of theorem 1.2.

\noindent Remark 1.2.1: Theorem 1.2 remains valid for semi-infinite matrices $L$;
the proof would only require minor modifications.

\section{Symmetries of the Toda lattice and the $w_2$-algebra}

Symmetries are $t$-dependent vector fields on the manifold of wave functions
$\Psi$, which commute with and are transversal to the Toda vector fields,
without affecting the $t$-variables. We shall need the following Lie algebra
splitting lemmas, dealing with operators and their eigenfunctions, due to
[A-S-V].

\proclaim Lemma 2.1.
Let $\DR$ be a Lie algebra with a vector space decomposition
$\DR=\DR_+\oplus\DR_-$ into two Lie subalgebras $\DR_+$ and $\DR_-$;
let $V$ be a representation space of $\DR$, and
let $\MR\subset V$ be a submanifold preserved under the vector fields
defined by the action of $\DR_-$,
i.e.,
$$
\DR_-\cdot x\subset T_x\MR,\quad\forall x\in\MR.
$$
For any function $p\colon\MR\to\DR$, let $\BY_p$ be the vector field
on $\MR$ defined by
$$
	\BY_p(x):= -p(x)_-\cdot x,\quad x\in\MR.
$$
(a) Consider a set $\AR$ of functions $p\colon\MR\to\DR$ such that
$$
	\BY_qp=[-q_-,p],\quad\forall p,q\in\AR,
$$
holds.
Then $\BY\colon p\mapsto\BY_p$ gives a Lie algebra homomorphism
of the Lie algebra generated by $\AR$ to the Lie algebra $\XR(\MR)$
of vector fields on $\MR$:
$$
	[\BY_{p_1},\BY_{p_2}]=\BY_{[p_1,p_2]},\quad\forall p_1,p_2\in\AR,
$$
and hence we can assume without loss of generality
that $\AR$ itself is a Lie algebra.\hfill\break
(b) Suppose for a subset ${\cal B}\subset\AR$ of functions
$$
	\BZ_q(x):=q(x)_+\cdot x\in T_x\MR,\quad\forall x\in\MR,q\in{\cal B},
$$
and hence defines another vector field $\BZ_q\in\XR(\MR)$ when $q\in{\cal B}$,
and such that $$
	\BZ_qp=[q_+,p],\quad\forall p\in\AR,q\in{\cal B},
$$
holds. Then
$$
	[\BY_p,\BZ_q]=0,\quad\forall p\in\AR,q\in{\cal B}.
$$

\medbreak
\underline{Remark 2.1.1}: A special case of this which applies to many
integrable systems is: $V=\DR'$, a Lie algebra containing $\DR$,
and $\DR$ acts on $\DR'$ by Lie bracket, i.e., $\BY_p(x)=[-p(x)_-,x]$,
etc.

\medbreak
{\sl Proof:} To sketch the proof, let $p_1$,~$p_2,~p\in\AR$ and $q\in {\cal B}$; then
 the commutators have the following form:
$$
[\BY_{p_1},\BY_{p_2}](x)=Z_1(x)~~\mbox{and}~~[\BZ_q,\BY_p](x)=Z_2x,
$$
where, using $ \DR_-$ and $\DR_+$ are Lie subalgebras,
\begin{eqnarray*}
Z_1&:=&\bigl(\BY_{p_1}(p_2)\bigr)_- -
\bigl(\BY_{p_2}(p_1)\bigr)_-+[p_{1-},p_{2-}]\\&=&[-p_{1-},p_2]_- -
[-p_{2-},p_1]_- + [p_{1-},p_{2-}]\\
&=&\bigl(-[p_{1-},p_2] - [p_1,p_{2-}] + [p_{1-},p_{2-}] -
		[p_{1+},p_{2+}]\bigr)_-=-[p_1,p_2]_-.\\
\end{eqnarray*}
and
\begin{eqnarray*}
	Z_2&:=&\bigl(\BZ_q(p)\bigr)_- + \bigl(\BY_p(q)\bigr)_+ - [q_+,p_-]\\
	&=&[q_+,p]_- + [-p_-,q]_+ - [q_+,p_-]\\
	&=&[q_+,p_-]_- + [-p_-,q_+]_+ - [q_+,p_-]
	=[q_+,p_-]-[q_+,p_-]=0,\\
\end{eqnarray*}
ending the proof of the lemma.

In the setup of the lemma, if we are given a Lie algebra (anti)homomorphism
$\phi:{\bf g}\to\AR$, we denote $\BY_{\phi(x)}$ by $\BY_x$ and $\BZ_{\phi(x)}$
by $\BZ_x$ if there is no fear of confusion.

\proclaim Theorem 2.2. Let $L$ represent an infinite symmetric tridiagonal
matrix, flowing according to the Toda vector fields. There is a Lie algebra
anti-homomorphism
$$
w_2^+=\{z^{n+\ell}(\frac{\pl}{\pl z})^n,n=0,\ell\in\BZ\mbox{ or }n=1,\ell\geq
-1\}\rg \left\{
\begin{tabular}{l}
vector fields on the\\
manifold of wave functions
\end{tabular}
\right\}$$
$$
z^{n+\ell}(\frac{\pl}{\pl z})^n\longrightarrow \BY_{\ell,n}\Psi=-
(M^nL^{n+\ell})_b\Psi
$$
satisfying
$$
[\BY_{\ell,0},\BY_{m,0}]=\frac{\ell}{2}\dt_{\ell+m},\quad
[\BY_{\ell,0},\BY_{m,1}]=\ell \BY_{m+\ell,0}\mbox{
and
}[\BY_{\ell,1},\BY_{m,1}]=(\ell-m)\BY_{m+\ell,1}.
$$ They commute with the Toda vector fields: $$
[\BY_{\ell,n},\frac{\pl}{\pl t_k}]=0.
$$
Note the vector fields $\BY_{\ell,n}$ induce vector fields on $S$ and $L=S\dt
S^{-1}$ $$
\BY_{\ell,n}(S)=-(M^nL^{n+\ell})_bS\mbox{  and 
}\BY_{\ell,n}(L)=[-(M^nL^{n+\ell})_b,L].
$$

\medbreak
{\sl Proof of Theorem 2.2:} Taking into account the notation of 1.13 and in
view of Lemma 2.1 and the remark 2.1.1, set
$$
\DR :=gl(\iy),\quad \DR_+:=\DR_s,\quad\DR_-:=\DR_b\quad\mbox{and}\quad
\DR':=\DR\times\DR
$$
on which $\DR$ acts via diagonal
embedding $\DR\hookrightarrow\DR':p\mapsto(p,p)$.
\begin{eqnarray*}
V	&:=&\DR\\
\MR	&:=&\mbox{respectively, the space of wave operators} S, \mbox{ of wave
functions } \Psi\\
& &\mbox{or of pairs }(L,M) =(S\dt
S^{-1},S(\vr+\frac{1}{2}\sum_1^{\iy}kt_k\dt^{k-1})S^{-1}),\mbox{ with an}\\
& &\mbox{infinite symmetric tridiagonal matrix } L\\
\AR&:=&\mbox{span }\left\{\begin{array}{lll}
M^nL^{n+\ell},& &n=0,\ell\in\BZ\\
\mbox{or }& &n=1,\ell\geq -1 
\end{array}
\right.\\
{\cal B}	&:=&\mbox{span }\{L^{\al},\al\in\BZ\}
\end{eqnarray*}
and
$$
{\bf g}:=w_2^+
$$
with the antihomomorphism $\phi :{\bf g}\to\AR$ given by 
$$
\phi(z^{\al}\pl_z^{\be}):=M^{\be} L^{\al}.
$$
Then the vector fields take the form:
$$\BY_p\Psi=-p_b\Psi,\quad\BY_pS=-p_bS,\quad p\in \AR$$
$$\BY_p(L,M)=([-p_b,L],[-p_b,M]),\quad p\in\AR$$
$$\begin{tabular}{lll}
$\BZ_{L^n/2}\Psi$&$=\frac{1}{2}(L^n)_s\Psi,\quad\quad
\BZ_{L^n/2}S$&$=-\frac{1}{2}(L^n)_bS$\\
        &$=\frac{\pl}{\pl t_n}\Psi$&$=\frac{\pl}{\pl t_n}S$
\end{tabular}
$$
\hfill by Theorem 1.2,
$$
\BZ_{L^n/2}(L,M)=([\frac{1}{2}(L^n)_s,L],[\frac{1}{2}(L^n)_s,M])=
\frac{\pl}{\pl t_n}(L,M)\quad\quad\mbox{by (1.18).}$$

Note that the vector fields
$$
\BY_{m,0}\equiv\BY_{L^m}\mbox{ all }m\in\BZ\mbox{ and
}\BY_{\ell,1}=\BY_{ML^{\ell+1}}, \mbox{ all }\ell\in\BZ,\geq -1
\leqno{(2.1)}
$$
are tangent to $\MR.$ Indeed for $m<0$, the vector field
reads $$\frac{\pl}{\pl s_m}\Psi=\BY_{m,0}\Psi=-(L^m)_b\Psi=-L^m\Psi=-z^m\Psi
\quad\mbox{for}\quad m<0.\leqno{(2.2)}
$$
The solution to this equation with initial condition $\Psi^{(0)}$ is given by
$$
\Psi=e^{-s_mz^m}\Psi^{(0)}(t,z)
$$
i.e., every component of the vector $\Psi^{(0)}$ is multiplied by the same exponential
factor, and so is each $\tau$-function:
$$
\tau_k(t)=\tau_k^0(t)e^{-ms_mt_{-m}}.
$$
Since the entries of the tridiagonal matrix only depend on ratios of
$\tau$-functions, this exponential factor is irrelevant for $L.$

In the same way $\BY_{m,0}$ ($m\geq 0$) is tangent to the space of symmetric
tridiagonal matrices, because the solution to
$$
\frac{\pl\Psi}{\pl s_m}=\BY_{m,0}\Psi=-(L^m)_b\Psi=(-L^m+(L^m)_s)\Psi
=-z^m\Psi+2\frac{\pl\Psi}{\pl t_m},\leqno{(2.3)}
$$
is given by
$$
\Psi=e^{\frac{1}{2}\Sg t_iz^i}\frac{\tau (t+2s-[z])}{\tau(t)}.
$$
Not only
the vector fields ${\Bbb Y}_{\ell ,0}$, but also the ${\Bbb
Y}_{\ell ,1}$'s ($\ell\geq -1$) are tangent to the space of symmetric
tridiagonal matrices, because \ALIGNER
{\Bbb Y}_{\ell ,1}(L) 
&=[-(ML^{\ell +1})_b,L]~,~~ \mbox{ having the form
}a_1\delta+\sum_0^{\iy}a_{-i}\delta^{-i}&\gauche{(2.4)}\\ &=[-ML^{\ell
+1},L]+[(ML^{\ell +1})_s,L]\\ &=L^{\ell +1}+[(ML^{\ell
+1})_s,L]=\mbox{ symmetric matrix for }\ell\geq -1.
 \aligner
With these data in mind, Lemma 2.1 implies Theorem 2.2.

\section{The action of the symmetries on $\tau$-functions}

The main purpose of this section is to show that the symmetry vector fields
$\BY$ defined on the manifold of wave function $\Psi$ induce certain precise 
vector fields on $\tau$, given  by the coefficients of the vertex operator
expansion. The precise statement is contained in theorem 3.2.  Before entering
these details, we need a general statement:

\proclaim Lemma 3.1. Any vector field $\BY$ defined on the manifold of wave
functions $\Psi$ and commuting with the Toda vector fields induce  vector fields
$\hat\BY$ on the manifold of $\tau$-functions; they are related as follows,
taking into account the fact that $\BY\log$ acts as a logarithmic
derivative:
\ALIGNER\BY \log \Psi_n &=
(e^{-\eta}- 1)~ \hat{\BY} \log \tau_n + \frac{1}{2}  \hat{\BY} \log
\frac{\tau_n}{\tau_{n+1}}&\gauche{(3.1)}\\  & \equiv \sum^{\iy}_{1} a_i^{(n)}
z^{-i} + a_0^{(n)}
\aligner
where $\hat\BY$ is a vector field acting
on $\tau$-functions; the part of (3.1) containing $e^{-\eta}-1$ is a
power series in $z^{-1}$ vanishing at $z=\iy$, whereas the other part is
$z$-independent. For any two vector fields $$ [\BY, \BY'] \log 
\Psi_n = (e^{-\eta}-1)~[\hat\BY,
\hat\BY'] \log \tau_n
+\frac{1}{2}[\hat\BY,
\hat\BY']\log
\frac{\tau_n}{\tau_{n+1}}, \leqno{(3.2)}
$$
showing that the map above from the algebra of vector fields on wave
functions to the algebra of vector fields on $\tau$-functions is homomorphism.

\medbreak
{\sl Proof:} In the computation below we use
$\ga_n=\sqrt{\frac{\tau_{n+1}}{\tau_n}}$ and the fact that $\BY:= ~^.$ commutes with
the Toda flows $\pl/\pl t_n$ and thus with
$\eta$: 
\ALIGNER 
(\log\Psi_n)^.&=(\log \frac{e^{-\eta}\tau_n}{\tau_n}-\log\ga_n)^.\quad\mbox{ see
(1.21)}&\gauche{(3.3)}\\
&=\frac{(e^{-\eta}\dot\tau_n)}{e^{-\eta}\tau_n}-\frac{\dot\tau_n}{\tau_n}+
\frac{1}{2}(\log\frac{\tau_n}{\tau_{n+1}})^.\\ 
&=(e^{-\eta}-1)(\log\tau_n)^.+\frac{1}{2}(\log\frac{\tau_n}{\tau_{n+1}})^.,
\quad\mbox{ using }[\eta,\BY]=0.
\aligner

\medbreak

Applying the second vector field $\BY'$ to relation (3.3) yields
$$
\BY'\BY\log\Psi_n=(e^{-\eta}-1)\BY'(\hat\BY\log\tau_n)+\frac{1}{2}\BY'(\hat\BY\log
\frac{\tau_n}{\tau_{n+1}})
$$
with
\begin{eqnarray*}
\BY'(\hat\BY\log f)&=&\BY'(\frac{\hat\BY f}{f})\\
&=&\frac{\hat\BY'(\hat\BY f)}{f}-\frac{(\hat\BY'
f)(\hat\BY f)}{f^2}\\
&=&\frac{\hat\BY'(\hat\BY f)}{f}-\frac{(\hat\BY' f)(\hat\BY f)}{f^2}.
\end{eqnarray*}
Using the relation above
twice leads at once to (3.2), ending the proof of Lemma 3.1.

The vector fields
$\BY_{\ell,n}$ on $\Psi$ induce certain
precise vector fields on $\tau$, constructed from
expressions $ W_{\ell}^{(n+1)}$ appearing in the vertex operator
expansion; notice the expressions $W_{\ell}^{(n+1)}$ differ
slightly from the customary ones, because of the
$1/2$ multiplying $t$ but not $\pl/\pl t$:
\ALIGNER
X(t,\lb,\mu)\tau
&=e^{\frac{1}{2}\sum_1^{\iy}t_i(\mu^i-\lb^i)}
e^{\sum_1^{\iy}(\lb^{-i}-\mu^{-i})\frac{1}{i}\frac{\pl}{\pl
t_i}}\tau&\gauche{(3.4)}\\
&=e^{\frac{1}{2}\Sg
t_i(\mu^i-\lb^i)}\tau(t+[\lb^{-1}]-[\mu^{-1}])\\
&=\sum_{k=0}^{\iy}\frac{(\mu -\lb)^k}{k!}\sum_{\ell
=-\iy}^{\iy}\lb^{-\ell -k} W_{\ell}^{(k)}(\tau).
\aligner
For instance
$$
W_n^{(0)} = J_n^{(0)} = \delta _{n,0}, \quad
\quad W_n^{(1)}=J_n^{(1)},\quad\quad 
W_n^{(2)}=J_n^{(2)}-(n+1)J_n^{(1)}\leqno{(3.5)} $$
with
$$
J_n^{(1)} =\left\{
\begin{tabular}{ll}
$\frac{\pl}{\pl t_n}$&for $n>0$\\
$0$&for $n=0$\\
$\frac{1}{2}(-n)t_{-n}$&for $n<0$
\end{tabular}
\right.\quad
J_n^{(2)} =\left\{
\begin{tabular}{ll}
$\renewcommand{\arraystretch}{0.5}
\begin{array}[t]{c}
\sum\\
{\scriptstyle i+j=n}\\
{\scriptstyle i,j\geq 1}
\end{array}
\renewcommand{\arraystretch}{1}
\frac{\pl^2}{\pl t_i\pl t_j}+
\renewcommand{\arraystretch}{0.5}
\begin{array}[t]{c}
\sum\\
{\scriptstyle -i+j=n}\\
{\scriptstyle i,j\geq 1}
\end{array}
\renewcommand{\arraystretch}{1}it_i\frac{\pl}{\pl t_j}$&for $n\geq 0$\\
$\frac{1}{4}\renewcommand{\arraystretch}{0.5}
\begin{array}[t]{c}
\sum\\
{\scriptstyle i+j=-n}\\
{\scriptstyle i,j\geq 1}
\end{array}
\renewcommand{\arraystretch}{1}(it_i)(jt_j)+
\renewcommand{\arraystretch}{0.5}
\begin{array}[t]{c}
\sum\\
{\scriptstyle i-j=-n}
\end{array}
\renewcommand{\arraystretch}{1}
it_i\frac{\pl}{\pl
t_j}$&for $n\leq 0.$ \end{tabular}
\right.
$$
We shall also need
$$
\LR^j_{\ell,1}:=W_{\ell}^{(2)}+2jW_{\ell}^{(1)}+(j^2-j)W_{\ell}^{(0)}=
J_{\ell}^{(2)}+(2j-\ell
-1)J_{\ell}^{(1)}+(j^2-j)J_{\ell}^{(0)}
\leqno{(3.6)} $$
$$
\LR^j_{\ell,0}:=2W_{\ell}^{(1)}+2jW_{\ell}^{(0)}=2J_{\ell}^{(1)}+2jJ_{\ell}^{(0)}.
$$
We also introduce $\tilde W_n^{(2)}$, which differs from $W_n^{(2)}$ above by
a factor $1/2$,
$$
\tilde W_n^{(2)}:=J_n^{(2)}-\frac{n+1}{2}J_n^{(1)},\mbox{ with
}\tilde W_n^{(2)}=W_n^{(2)}=J_n^{(2)}\mbox{ for }n=-1,0,\leqno{(3.7)} $$
and an operator $B_m$ in $z$ and $t$
$$
B_m:=-z^{m+1}\frac{\pl}{\pl z}+\sum_{n>\max(-m,0)}nt_n\frac{\pl}{\pl
t_{n+m}},\quad m\in\BZ,\leqno{(3.8)} $$
which restricted to functions $f(t_1,t_2,\pp)$ of $t\in\BC^{\iy}$ only yields
$$
B_mf=J_m^{(2)}f\mbox{  for  }m=-1,0,1.\leqno{(3.9)}
$$
The expressions $J_n^{(2)}$ form a Virasoro algebra with central extension
$$
[J_{\ell}^{(1)},J_m^{(1)}]=\frac{\ell}{2}\dt_{\ell +m}\quad
[J_{\ell}^{(1)},J_m^{(2)}]=\ell J_{m+\ell}^{(1)}\leqno{(3.10)}
$$
$$
[J_{\ell}^{(2)},J_m^{(2)}]=(\ell -m)J_{\ell
+m}^{(2)}+\frac{\ell^3-\ell}{12}\dt_{\ell +m}
$$
and consequently the following Virasoro commutation relations hold, upon
setting 
$$ 
V_{\ell} = J_{\ell}^{(2)} + (a\ell + b) J_{\ell}^{(1)} \quad ,\quad c_{\ell}
= \frac{\ell^3-\ell}{12} + \frac{\ell(b^2-a^2\ell^2)}{2}, \leqno{(3.11)} $$ 
$$
[V_{\ell},V_{m}] = (l-m)V_{\ell+m} + c_{\ell}\delta_{\ell+m,0}.
$$ 
In particular, observe
$$
[W_{\ell}^{(1)},W_m^{(2)}]=\ell W_{\ell +m}^{(1)}+\frac{\ell^2-\ell}{2}\dt_{\ell
+m},\quad [W_{\ell}^{(2)},W_m^{(2)}]=(\ell -m)W_{\ell
+m}^{(2)}-5\frac{(\ell^3-\ell)}{12}\dt_{\ell +m}$$
and\footnote{with $c_{\ell,j}=\ell(\ell
+2j-1)\quad c'_{\ell,j}=-\frac{\ell}{12}(5\ell^2+24j^2+7)$}
$$
[\LR_{\ell,0}^{(j)},\LR_{m,1}^{(j)}]=\ell \LR_{\ell
+m,0}^{(j)}+c_{\ell,j}\dt_{\ell
+m},~~[\LR_{\ell,1}^{(j)},\LR_{m,1}^{(j)}]=(\ell -m)\LR_{\ell
+m,1}^{(j)}+c'_{\ell,j}\dt_{\ell +m}.\leqno{(3.12)}
$$

\bigbreak

The purpose of this section is to prove the following relationship between
the action of the symmetries on $\Psi$ and $\tau$.

\proclaim Fundamental Theorem 3.2. The
following relationship holds  
$$
\BY_{\ell,n}\log\Psi
 ={(e^{-\eta}-1)\LR_{\ell,n}\log\tau + \frac{1}{2}
\Bigl(
\LR_{\ell,n}\log\tau-\bigl(\LR_{\ell,n}\log\tau\bigr)_{\delta}\Bigr)}\leqno{(3.13)}
$$
$$
\begin{tabular}{lll}
\hspace{8cm} for&$n=0$,&all $\ell\in\BZ$\\
   &$n=1$,&all $\ell \geq -1$
\end{tabular}
$$
where $\BY_{\ell,n}\log$ and $\LR_{\ell,n}\log$ act as logarithmic derivatives and
where
$$
(\tau_{\delta})_j=\tau_{j+1}~~~~~
(\LR_{\ell,n}\tau)_j=\LR_{\ell,n}^j\tau_j~~~~
\Bigl((\LR_{\ell,n})_{\delta}\Bigr)_j=\LR_{\ell,n}^{j+1}.
$$

\proclaim Corollary 3.2.1. The following holds for $m\geq -1:$
 $$
-\Bigl(M L^{m+1}+\frac{m+1}{2}
L^m\Bigr)_b\Psi=\Bigl(z^j\Psi_j((e^{-\eta}-1)
\frac{(J^{(2)}_m+2jJ_m^{(1)}+j^2 J^{(0)}_m)\tau_j}{\tau_j}
$$
$$
+\frac{1}{2}\Bigl(\frac{(J^{(2)}_m+2jJ_m^{(1)}+j^2J^{(0)}_m)\tau_j}{\tau_j}-
\frac{(J^{(2)}_m+2(j+1)J_m^{(1)}+(j+1)^2J^{(0)}_m)\tau_{j+1}}{\tau_{j+1}}
)\Bigr)\Bigr)_{j\in\BZ}, $$
where $J_{\ell}^{(i)}$ is defined in (3.5).

Remark that the statements of the theorem and the corollary are equally valid for
semi-infinite matrices. Before giving the proof we need three lemmas.

\proclaim Lemma 3.3. The operator $B_m$ defined in (3.8) interacts with $\eta
=\sum_1^{\iy}\frac{z^{-i}}{i}\frac{\pl}{\pl t_i}$ and $\Sg =\sum_1^{\iy}t_iz^i$
as follows\footnote{with the understanding that $$
\sum^{\al}_{j=1}=0\quad\mbox{if}\quad\al <1.
$$}
$$
[B_m,\eta]=\sum^m_{k=1}z^{m-k}\frac{\pl}{\pl t_k}\quad\mbox{and}\quad
B_m\Sg=-\sum^{-m}_{k=1}z^{m+k}kt_k\leqno{(3.14)}
$$
$$
[B_m,e^{-\eta}]=-e^{-\eta}[B_m,\eta];
$$
thus
$$
[B_m,\eta]=0\mbox{ when }m\leq 0\mbox{ and }B_m\Sg =0\mbox{ when }m\geq 0.
$$
Moreover given an arbitrary function $f(t_1,t_2,\pp)$ of $t\in\BC^{\iy}$, define
$$
\Phi =e^{\Sg/2}\frac{e^{-\eta}f}{f}.\leqno{(3.15)}
$$
Then we have for $m\geq 0$
$$
\begin{tabular}{lll}
(i)& & $-\frac{1}{2}z^{-m}\Phi
=\Phi(e^{-\eta}-1)\frac{W^{(1)}_{-m}(f)}{f},\quad m>0$\\
(ii)&  &$(B_{-m}+\frac{mt_m}{2}) \Phi =\Phi (e^{-\eta}-1) \frac{ 
\tilde W^{(2)}_{-m}(f)}{f},$\\
(iii)& &$B_m\Phi=\Phi\Bigl((e^{-\eta}-1)\frac{(B_m-[B_m,\eta])f}{f}-
\frac{[B_m,\eta]f}{f}\Bigr).$
\end{tabular}\leqno{(3.16)}
$$

\medbreak
{\sl Proof:} The first commutation relation (3.14) follows from a
straightforward computation:
\begin{eqnarray*}
[B_m,\eta]&=&\Bigl[-z^{m+1}\frac{\pl}{\pl z},\eta\Bigr]+\Bigl[
\sum_{k>\max(-m,0)}kt_k\frac{\pl}{\pl t_{k+m}},\eta\Bigr]\\
&=&\sum_1^{\iy}z^{m-i}\frac{\pl}{\pl t_i}-\sum_1^{\iy}z^{-j}\frac{\pl}{\pl
t_{j+m}} = \sum_1^m z^{m-i}\frac{\pl}{\pl t_i}.
\end{eqnarray*}
The third commutation relation (3.14) follows at once from the fact that the
bracket $[B_m,.]$ is a derivation and that $\left[[B_m,\eta],\eta \right]$. Also
$$
B_m\Sg =-\sum_1^{\iy}kt_kz^{k+m}+\sum_{k>\max(-m,0)}kt_k \,z^{k+m}=-
\sum_1^{\max(-m,0)}kt_k\,z^{k+m}.
$$
Next, given $\Phi$ defined as (3.15), we have that (using $B_m$ is a
derivation and (3.14)), \medbreak
\noindent (3.17)
\begin{eqnarray*}
B_m\Phi&=&B_me^{\Sg/2}\frac{e^{-\eta}f}{f}\\
&=&\frac{e^{-\eta}f}{f}B_me^{\Sg/2}+e^{\Sg/2}B_m\frac{e^{-\eta}f}{f}\\
&=&\Phi\Bigl((e^{-\eta}-1)\frac{B_mf}{f}+\frac{1}{2}B_m\Sg-e^{-\eta}
\frac{[B_m,\eta]f}{f}\Bigr)\\
&=&\Phi\Bigl((e^{-\eta}-1)\frac{(B_m-[B_m,\eta])f}{f}+\frac{1}{2}B_m\Sg -
\frac{[B_m,\eta]f}{f}\Bigr)
\end{eqnarray*}
which yields (3.16) (iii) for $m\geq 0$, taking into account the fact that
$B_m\Sg =0$ for $m\geq 0.$
Equation (3.16) (i) is straightforward, using $W^{(1)}_{-m}=\frac{1}{2}mt_m.$

Proving (3.16) (ii) is a bit more involved; indeed first observe that for
$m\geq 0$ \begin{eqnarray*}
\tilde W_{-m}^{(2)}f
&=&\Bigl(\sum_{n>m}nt_n\frac{\pl}{\pl
t_{n-m}}+\frac{1}{4}\sum_{n=1}^{m-1}nt_n(m-n)t_{m-n}
+\frac{1}{4}(m-1)mt_m\Bigr)f\\ 
&=&B_{-m}f
+\frac{1}{4}\sum_{n=1}^{m-1}nt_n(m-n)t_{m-n}f +\frac{1}{4}(m-1)mt_mf;
\end{eqnarray*}
therefore
$$(e^{-\eta}-1)\frac{B_{-m}(f)}{f}\leqno{(3.19)}$$
\begin{eqnarray*}
&=&(e^{-\eta}-1)\Bigl(\frac{ 
\tilde W_{-m}^{(2)}f}{f}-\frac{1}{4}
\sum_{n=1}^{m-1}nt_n(m-n)t_{m-n}
-\frac{1}{4}m(m-1)t_m \Bigr)\\
&=&(e^{-\eta}-1)\frac{\tilde W_{-m}^{(2)}(f)}{f}\\ &
&-\frac{1}{4}\sum_{n=1}^{m-1}n(m-n)
\bigl((t_n-\frac{1}{n}z^{-n})(t_{m-n}-\frac{1}{m-n}
z^{-m+n})-t_nt_{m-n}\bigr)\\  & &-\frac{1}{4}m(m-1)
\bigl((t_m-\frac{1}{m}z^{-m})-t_m\bigr)\\
&=&(e^{-\eta}-1)\frac{\tilde W_{-m}^{(2)}(f)}{f}
+\frac{1}{2}\sum_{n=1}^{m-1}nt_nz^{n-m}. 
\end{eqnarray*}
Using expression (3.17) for $B_m\Phi$, $-m\leq 0$, and $[B_m,\eta]=0$ when
$m\leq 0$, we find \begin{eqnarray*}
(B_{-m}+\frac{mt_m}{2})\Phi&=&\Phi\Bigl((e^{-\eta}-1)\frac{B_{-m}f}{f}+
\frac{1}{2}B_{-m}\Sg+\frac{mt_m}{2}\Bigr)\\
&=&\Phi\Bigl((e^{-\eta}-1)\frac{B_{-m}f}{f}-
\frac{1}{2}\sum^{m-1}_{k=1}z^{-m+k}kt_k\Bigr),\mbox{ using (3.14)}\\
&=&\Phi(e^{-\eta}-1)\frac{\tilde W_{-m}^{(2)}f}{f},\mbox{ using (3.19),}
\end{eqnarray*}
ending the proof of (3.14) (ii) and thus of Lemma 3.3.

\proclaim Lemma 3.4. The following holds $(\nu \equiv \vr\pl = \mbox{ diag
}(\cdots,\nu_i=i,\cdots))$
$$
(PL^2)_s=\nu L_s+L_-\leqno{\rm (i)}
$$
$$
\frac{(PL^2)_s\Psi}{\Psi}=(e^{-\eta}-1)(2\nu -I)\frac{\pl}{\pl
t_1}\log\tau +\nu\Bigl(z-\frac{\pl}{\pl
t_1}\log\frac{\tau_s}{\tau}\Bigr).\leqno{\rm (ii)} $$

\medbreak
{\sl Proof:} First consider
$$
PL^{m+1}=S\vr\dt^{m+1}S^{-1}=S\nu\dt^mS^{-1}=[S,\nu]\dt^mS^{-1}+\nu
S\dt^mS^{-1}=[S,\nu]\dt^mS^{-1}+\nu L^m.\leqno{(3.20)} $$
Since $\nu=\vr\dt$ is diagonal, since $S\in\DR_{-\iy,0}$ and so $[S,\nu]\in
\DR_{-\iy,-1}$, we have 
$$
[S,\nu]\dt S^{-1}\in \DR_{-\iy,0}\quad\mbox{and thus}\quad ([S,\nu]\dt 
S^{-1})_s=0.
$$
This combined with the above observation leads to\footnote{$L_{++}$ denotes the
strictly uppertriangular part of $L$, i.e. the projection of $L$ onto
$\DR_{1,\iy}$}
\ALIGNER
(PL^2)_s=(\nu L)_s&=\nu L_{++}-(\nu L_{++})^{\top}&\gauche{(3.21)}\\
&=\nu L_{++}-L_-\nu\\
&=\nu(L_{++}-L_-)+\nu L_--L_-\nu\\
&=\nu L_s+[\nu,L_-]\\
&=\nu L_s+L_-, \quad\mbox{since } [\nu,L_-]=L_-,
\aligner
since $L_-$ has one subdiagonal.  In order to evaluate 
$(PL^2)_s\Psi$, we need to know $L_s\Psi$ and $L_-\Psi.$ Anticipating (3.25),
we have by (2.3) and (3.3)\ALIGNER
\BY_{1,0}\Psi=-L_b\Psi&=2\Psi(e^{-\eta}-1)\frac{\pl}{\pl t_1}\log\tau-\Psi
\frac{\pl}{\pl t_1}\log\frac{\tau_{\dt}}{\tau}\\
&=2\Psi(e^{-\eta}-1)\frac{\pl}{\pl t_1}\log\tau-L_0\Psi, &\mbox{by}~ (1.16),
\aligner
and, since $L_b=2L_-+L_0$,
$$
L_-\Psi=\frac{1}{2}(L_b-L_0)\Psi=-\Psi(e^{-\eta}-1)
\frac{\pl}{\pl t_1}\log\tau,\leqno{(3.22)} $$
whereas, using
$$
\Psi=e^{\frac{1}{2}\Sg
t_iz^i}\Bigl(z^j\ga_j^{-1}\frac{\tau_j(t-[z^{-1}])}{\tau_j(t)}\Bigr), $$
we have (using the logarithmic derivative)
$$
L_s\Psi=2\frac{\pl\Psi}{\pl t_1}=2\Psi(e^{-\eta}-1)\frac{\pl}{\pl
t_1}\log\tau+z\Psi-\Psi\frac{\pl}{\pl
t_1}\log\frac{\tau_{\dt}}{\tau}.\leqno{(3.23)} $$ Using these two formulas
(3.22) and (3.23), in (3.21) we have \begin{eqnarray*}
(PL^2)_s\Psi&=&\nu L_s\Psi+L_-\Psi\\
&=&\Psi(e^{-\eta}-1)(2\nu -I)\frac{\pl}{\pl t_1}\log\tau+z\nu\Psi-\nu\Psi\frac{\pl}{\pl t_1}
\log\frac{\tau_{\dt}}{\tau},
\end{eqnarray*}
ending the proof of Lemma 3.4.

This lemma proves the main statement about symmetries for the $s\ell(2,\Bbb
C)$ part of the Virasoro symmetry algebra; this is the heart of the matter.

\proclaim Lemma 3.5. The vector fields  $\{\BY_{-1,1},\BY_{0,1},\BY_{1,1}\}$
form a representation of $s\ell(2,\BC)$ and induce vector fields on $\tau$ as
follows (in the notation (3.6)):
 $$
\BY_{\ell,1}\log\Psi=\frac{-(ML^{\ell+1})_b\Psi}{\Psi}=(e^{-\eta}-1)
\frac{\LR_{\ell,1}\tau}{\tau}+\frac{1}{2}\left(\frac{\LR_{\ell,1}\tau}{\tau}-
(\frac{\LR_{\ell,1}\tau}{\tau})_{\dt}\right)\leqno{(3.24)}
$$
\hfill for $\ell=-1,0,1.$\newline Also
$$
\BY_{\ell,0}\log\Psi=\frac{-(L^{\ell})_b\Psi}{\Psi}=(e^{-\eta}-1)
\frac{\LR_{\ell,0}\tau}{\tau}+\frac{1}{2}\left(\frac{\LR_{\ell,0}\tau}{\tau}-
(\frac{\LR_{\ell,0}\tau}{\tau})_{\dt}\right)\leqno{(3.25)}
$$
\hfill for $\ell\in\BZ.$

\medbreak
{\sl Proof of Lemma 3.5:} Relation (3.25) will first be established for $ \ell
=-m<0:$
\begin{eqnarray*}
\BY_{-m,0}\Psi
&=&-(  L^{-m})_b  
\Psi=-L^{-m}\Psi,\quad\mbox{since}\quad L^{-m}\in\DR_b
\quad\mbox{for}\quad m\geq
0\\ &=&-z^{-m} \Psi \\ 
&=&2\ga^{-1}\chi \left(-\frac{1}{2}z^{-m}(e^{\Sigma
/2}\frac{e^{-\eta}\tau_n}{\tau_n})\right)_{n\in \BZ}\\ &=& 2\Psi (e^{-\eta} -1)
\frac{ W_{-m}^{(1)} \tau}{\tau},~~\mbox{applying (3.16)(i) componentwise to}\\
& &\quad\quad\quad\quad\quad\quad\quad\quad\quad\quad\quad
\Phi=e^{\Sg/2}\frac{e^{-\eta}\tau_n}{\tau_n}\\
&=&\Psi \Bigl((e^{-\eta}-1) \frac{2W_{-m}^{(1)}\tau}{\tau}+
\frac{1}{2}\bigl(\frac{2W_{-m}^{(1)}\tau}{\tau}-(\frac{
2W_{-m}^{(1)}\tau}{\tau})_{\delta}\bigr)\Bigr),
\end{eqnarray*}
the difference in brackets vanishing identically.
The same will now be established for $\ell=m>0;$ indeed
\begin{eqnarray*}
\BY_{m,0}\Psi &\equiv&-(L^m)_b \Psi\\ 
&=&(-L^m + (L^m)_s)\Psi\\
&=&-z^m\Psi +2\frac{\pl}{\pl
t_m} \Psi~,~~~\mbox{  using (1.16)}\\
&=&-z^m\Psi+z^m\Psi + 2e^{\Sg/2} \chi
\frac{\pl}{\pl
t_m}(\frac{e^{-\eta}\tau}{\tau}\ga^{-1}),\\
& &\quad\quad\quad\quad\quad\quad\mbox{      using
}\Psi=e^{\Sg/2}\frac{e^{-\eta}\tau}{\tau}\ga^{-1} \chi\\
&=&2e^{\Sg/2} \chi \ga^{-1}\frac{e^{-\eta}\tau}{\tau}
\Bigl((e^{-\eta}-1)\frac{\pl}{\pl t_m} \log \tau +
\ga\frac{\pl}{\pl t_m} \ga^{-1} \Bigr),\quad\mbox{using (3.3)}\\
&=& \Psi \Bigl((e^{-\eta}-1)\frac{2 W_m^{(1)}\tau}{\tau} +
\frac{1}{2} (\frac{2W_m^{(1)}\tau}{\tau}-(\frac{
2W_m^{(1)}\tau}{\tau} )_{\delta})\Bigr)\\
& &\mbox{ since
}\ga_j = \sqrt{\frac{\tau_{j+1}}{\tau_j}}, \mbox{ and
}\ga
\frac{\pl}{\pl t_m} \ga^{-1}= -\frac{\pl}{\pl t_m }\log
\ga.
\end{eqnarray*}
Relation (3.25) for $\ell=0$ is obvious, since 
$$\frac{\BY_{0,0}\Psi}{\Psi}=\frac{-(L^0)_b \Psi}{\Psi}=-I=\frac{1}{2}
(2\nu-2(\nu)_{\delta})= \frac{1}{2}\left(\frac{2\nu \tau}{\tau}-(\frac{2\nu
\tau}{\tau})_{\delta}\right)=\frac{1}{2}\left(\frac{\LR_{0,0}\tau}{\tau}-
(\frac{\LR_{0,0}\tau}{\tau})_{\dt}\right)$$

To prove (3.24), consider now
\ALIGNER
(3.26)~&-(ML^{m+1})_b\Psi&\\
&=(-ML^{m+1}+(ML^{m+1})_s)\Psi\\
&=\Bigl(-z^{m+1}\frac{\pl}{\pl z}+\frac{1}{2}\sum_1^{\iy}kt_k(L^{k+m})_s+(
PL^{m+1})_s\Bigr)\Psi\\
&=\Bigl(-z^{m+1}\frac{\pl}{\pl z}+\frac{1}{2}\sum_{k>\max(-m,0)}kt_k(L^{k+m})_s+(
PL^{m+1})_s\Bigr)\Psi, \quad \quad \mbox{using (1.28)}\\
&\hspace{6cm}\mbox{since }(L^{\al})_s=0\mbox{ for }\al\leq 0\\
&=\Bigl(-z^{m+1}\frac{\pl}{\pl z}+\sum_{k>\max(-m,0)}kt_k\frac{\pl}{\pl
t_{k+m}}\Bigr)\Psi+(PL^{m+1})_s\Psi\\
&=B_m\Psi+(PL^{m+1})_s\Psi\quad\quad\mbox{using the definition (3.8) of
}B_m,\\ &=(z^j\ga_j^{-1}B_m
(e^{\Sigma/2}\frac{e^{-\eta}\tau_j}{\tau_j}))_{j\in\BZ}-z^m\nu\Psi-\Psi
B_m\log\ga+(PL^{m+1})_s\Psi,
\aligner
remembering the definition (1.21) of $\Psi$, using the definition (3.8)
of $B_m$, and using the fact that $B_m$ is a derivation. 
\bigbreak

\noindent 1. For $m=-1,0$, we have
$$
PL^{m+1}\in\DR_{-\iy,0}\quad\mbox{ and thus }\quad (PL^{m+1})_s=0.
$$
We set $m\mapsto -m$ with $m\geq 0.$
Componentwise, the above expression reads, by adding and subtracting $mt_m/2$
($m\geq 0$) and using definition (3.8) of $B_{-m}$ and $\nu =
\mbox{diag}(\cdots, i,\cdots)$: \medbreak \noindent $(-(ML^{-m+1})_b\Psi)_j$
\begin{eqnarray*}
&=&z^j\ga_j^{-1}\left(B_{-m}+ \frac{mt_m}{2} - j z^{-m}\right)
e^{\Sigma/2}\frac{e^{-\eta}\tau_j}{\tau_j}
-z^j \Psi_j \left(
B_{-m} (\log \ga_j)+ 
\frac{mt_m}{2}\right)\\ 
&=&z^j\ga_j^{-1}\left(B_{-m}+\frac{mt_m}{2}+2j(1-\delta_{0,m})(-\frac{z^{-m}}{2})\right)
(e^{\Sigma/2}\frac{e^{-\eta}\tau_j}{\tau_j})\\
& &-z^j
\Psi_j\left(
B_{-m}\log\ga_j+\frac{mt_m}{2}\right)-\frac{1}{2}z^j\Psi_j 2j\delta_{0,m}\\ 
&=&
z^j\Psi_j (e^{-\eta}-1)\bigl(\frac{\tilde W^{(2)}_{-m}(\tau_j)}{\tau_j} + 2j
\frac{ W^{(1)}_{-m}(\tau_j)}{\tau_j}\bigr)~~\mbox{  (using (3.16)(i) and
(ii))}\\  
& &+\frac{1}{2} z^j \Psi_j \Bigl( -\frac{ \tilde W^{(2)}_{-m}(\tau_{j+1})}
{\tau_{j+1}}+ \frac{ \tilde W^{(2)}_{-m}(\tau_j)}{\tau_j} \Bigr)~~\mbox{ 
(using }\ga_j=\sqrt{\frac{\tau_{j+1}}{\tau_j}} \mbox{, (3.9) and (3.7))}\\ 
& &+\frac{1}{2} z^j \Psi_j \left(-2(j+1) \frac{ 
W^{(1)}_{-m}(\tau_{j+1})}{\tau_{j+1}} +2j\frac{ 
W^{(1)}_{-m}(\tau_j)}{\tau_j} \right) ~\mbox{  (using
}\frac{mt_m}{2}=W^{(1)}_{-m})\\
& & +\frac{1}{2} z^j\Psi_j(-2j\delta_{0,m})\\
&=& z^j\Psi_j \Bigl( (e^{-\eta}-1)
\frac{( \tilde W_{-m}^{(2)} + 2j 
W_{-m}^{(1)}+(j^2-j)W_{-m}^{(0)})\tau_j}{\tau_j}\\ 
& & +\frac{1}{2}\Bigl(\frac{(
\tilde W_{-m}^{(2)} + 2j 
W_{-m}^{(1)}+(j^2-j)W_{-m}^{(0)})\tau_j}{\tau_j} \\
& & \quad\quad -\frac{( \tilde
W_{-m}^{(2)} + 2(j+1) 
W_{-m}^{(1)}+((j+1)^2-(j+1))W_{-m}^{(0)})\tau_{j+1}}{\tau_{j+1}}\Bigr) 
\Bigr),
\end{eqnarray*} 
since, in the last line, $-2j=(j^2-j)-\left((j+1)^2-(j+1)\right)$ and
$\delta_{0,m}=W^{(0)}_{-m}$. Thus relation (3.24) for
$m=-1$ and
$0$ follows from the simple observation (3.7) that 
$\tilde W_{-m}^{(2)}=W_{-m}^{(2)}$ for $m=0,1 $.

\bigbreak

\noindent 2. For $m=1$ in (3.24), we use the identities in Lemma 3.3(iii) and
Lemma 3.4(ii) in (3.26):
\medbreak
\noindent $-(ML^2)_b\Psi$
\begin{eqnarray*}
&=&\Psi(e^{-\eta}-1)\frac{(B_1-[B_1,\eta])\tau}{\tau}-\Psi
\frac{[B_1,\eta]\tau}{\tau}-z\nu\Psi-\frac{1}{2}\Psi
B_1\log\frac{\tau_{\dt}}{\tau}+(PL^2)_s\Psi\\
& &\hspace{9cm}\mbox{using (3.16)(iii)}\\
&=&\Psi(e^{-\eta}-1)(J_1^{(2)}-\frac{\pl}{\pl t_1})\log\tau-\Psi
\frac{\pl}{\pl t_1}\log\tau-z\nu\Psi-\frac{1}{2}\Psi
J_1^{(2)}\log\frac{\tau_{\dt}}{\tau},\\
& &\quad\quad +\Psi(e^{-\eta}-1)(2\nu-I)\frac{\pl}{\pl
t_1}\log\tau+z\nu\Psi-\nu\Psi\frac{\pl}{\pl t_1}\log\frac{\tau_{\dt}}{\tau}\\
&\,&\hspace{2cm}\mbox{using (3.9), }[B_1,\eta]=\frac{\pl}{\pl t_1}\mbox{ (see
(3.14)), and Lemma 3.4(ii), }\\
&=&\Psi\Bigl((e^{-\eta}-1)(J_1^{(2)}+(2\nu-2)\frac{\pl}{\pl t_1})\log\tau
+\frac{1}{2}(J_1^{(2)}+(2\nu-2)\frac{\pl}{\pl t_1})\log\tau\\
& &\hspace{2cm}-\frac{1}{2}(J_1^{(2)}+2\nu\frac{\pl}{\pl t_1})\log\tau_{\dt}
\Bigr),\\
&=& \Psi\left((e^{-\eta}-1)
\frac{\LR_{1,1}\tau}{\tau}+\frac{1}{2}\left(\frac{\LR_{1,1}\tau}{\tau}-
(\frac{\LR_{1,1}\tau}{\tau})_{\dt}\right)\right),\\
\end{eqnarray*}
using in the last line, the definition (3.6) of $\LR_{1,1}$ and
$\LR^{j+1}_{1,1}\tau_{j+1}=((\LR_{1,1}\tau)_{\delta})_j$, establishing (3.24)
for $m=1$, thus ending the proof of Lemma 3.5.

\medbreak

{\sl Proof of Theorem 3.2:} The only remaining point is to establish (3.13) for
$n=1$ and all $\ell\geq 2.$ To do this we use the underlying Lie algebra
structure. First we have the following identity, using (3.2) and
(3.1), $$
(e^{-\eta}-1)\Bigl[\hat\BY_{z^{\al}(\frac{\pl}{\pl
z})^{\beta}},\hat\BY_{z^{\al'}(\frac{\pl}{\pl
z})^{\beta'}}\Bigr]\log\tau_n+\frac{1}{2}
\Bigl[\hat\BY_{z^{\al}(\frac{\pl}{\pl
z})^{\beta}},\hat\BY_{z^{\al'}(\frac{\pl}{\pl
z})^{\beta'}}\Bigr]\log
\frac{\tau_n}{\tau_{n+1}} $$
\begin{eqnarray*}
&=&\Bigl[\BY_{z^{\al}
(\frac{\pl}{\pl z})^{\be}},\BY_{z^{\al'}
(\frac{\pl}{\pl z})^{\be'}}\Bigr]\log\Psi_n\\
&=&\BY_{\Bigl[z^{\al'}
(\frac{\pl}{\pl z})^{\be'},z^{\al}(\frac{\pl}{\pl
z})^{\be}\Bigr]}\log\Psi_n\\ &=&(e^{-\eta}-1)\hat\BY_{\Bigl[z^{\al'}
(\frac{\pl}{\pl z})^{\be'},z^{\al}(\frac{\pl}{\pl z})^{\be}\Bigr]}
\log\tau_n+\frac{1}{2}
\hat\BY_{\Bigl[z^{\al'}
(\frac{\pl}{\pl z})^{\be'},z^{\al}(\frac{\pl}{\pl
z})^{\be}\Bigr]}\log\frac{\tau_n}{\tau_{n+1}}. \end{eqnarray*}
The terms containing $(e^{-\eta}-1)$ in the first and last
expressions are power series in $z^{-1}$, with no constant term; the second terms are
independent of
$z.$ Therefore, equating constant terms yield:
$$
\Bigl(\Bigl[\hat\BY_{z^{\al}(\frac{\pl}{\pl
z})^{\beta}},\hat\BY_{z^{\al'}(\frac{\pl}{\pl
z})^{\beta'}}\Bigr]-
\hat\BY_{\Bigl[z^{\al'}(\frac{\pl}{\pl
z})^{\be'},z^{\al}(\frac{\pl}{\pl
z})^{\be}\Bigr]}\Bigl)\log\frac{\tau_n}{\tau_{n+1}}=0\leqno{(i)}$$
and thus also
$$
 (e^{-\eta}-1)\Bigl(\Bigl[\hat\BY_{z^{\al}(\frac{\pl}{\pl
z})^{\beta}},\hat\BY_{z^{\al'}(\frac{\pl}{\pl
z})^{\beta'}}\Bigr]-\hat\BY_{\Bigl[z^{\al'}(\frac{\pl}{\pl
z})^{\be'},z^{\al}(\frac{\pl}{\pl
z})^{\be}\Bigr]}\Bigl)\log\tau_n=0.
$$
Since $(e^{-\eta}-1)f=0$ implies
$f=$ constant, there exists a constant $c$ depending on
$\al,\be,\al',\be'$ and $n$ such that 
$$
 \Bigl(\Bigl[\hat\BY_{z^{\al}(\frac{\pl}{\pl
z})^{\beta}},\hat\BY_{z^{\al'}(\frac{\pl}{\pl
z})^{\beta'}}\Bigr]-\hat\BY_{\Bigl[z^{\al'}(\frac{\pl}{\pl
z})^{\be'},z^{\al}(\frac{\pl}{\pl
z})^{\be}\Bigr]}\Bigl)\log\tau_n=c_{\al\be\al'\be',n};
\leqno{(ii)}
$$
relation (i) says $c_{\al\be\al'\be'}$ is independent of $n$; hence (ii) reads

The two relations (i) and (ii) combined imply
$$
\Bigl(\Bigl[\hat\BY_{z^{\al}(\frac{\pl}{\pl
z})^{\beta}},\hat\BY_{z^{\al'}(\frac{\pl}{\pl
z})^{\beta'}}\Bigr]-
\hat\BY_{\Bigl[z^{\al'}(\frac{\pl}{\pl
z})^{\be'},z^{\al}(\frac{\pl}{\pl
z})^{\be}\Bigr]}-c_{\al\be\al'\be'}\Bigl)\tau_n=0
\leqno{(3.27)}
$$
with
$$
c_{\al\be\al'\be'}\quad\mbox{independent of }n.
$$
Applying (3.27) to
$$
\Bigl[z^{m+1}\frac{\pl}{\pl z},z^{\ell}\Bigr]=\ell
z^{m+\ell}\quad\mbox{and}\quad\Bigl[z^{\ell+1}\frac{\pl}{\pl
z},z^{m+1}\frac{\pl}{\pl z}\Bigr]= (m-\ell)z^{m+\ell+1}\frac{\pl}{\pl
z}, $$ leads to
$$
\Bigl(\Bigl[\hat\BY_{\ell,0},\hat\BY_{m,1}\Bigr]-\ell 
\hat\BY_{\ell+m,0}-c_{\ell,m}\Bigl)\tau_n=0
\leqno{(3.28)}
$$
and
$$
\Bigl(\Bigl[\hat\BY_{m,1},\hat\BY_{\ell,1}\Bigr]-(m-\ell) 
\hat\BY_{m+\ell,1}-c'_{m,\ell}\Bigl)\tau_n=0.
$$
By virtue of (3.12), we have
$$
[\LR_{\ell,0},\LR_{m,1}]-\ell 
\LR_{m+\ell,0}-\mbox{ constant }=0.
\leqno{(3.29)}
$$
According to (3.25) we have
$$
\hat\BY_{\ell,0}=\LR_{\ell,0}
$$
implying by subtracting (3.29) from (3.28)
$$
[\LR_{\ell,0},\hat\BY_{m,1}-\LR_{m,1}]=\mbox{ constant, for }\ell,m\in\BZ, m\geq
-1.$$
The only operator commuting (modulo constant) with all
$\LR_{\ell,0}=2\frac{\pl}{\pl t_1}+(-\ell)t_{-\ell}$ is given by linear
combinations of a constant, $t_{\al}$ and $\pl/\pl t_{\al}$, i.e.:
\ALIGNER
\tilde\BY_{m,1}-\LR_{m,1}&=\sum_{j=-\iy}^{\iy}c_j^{(m)}
J^{(1)}_{j+m}, \mbox{ for }m\geq 2,~~(c^{(m)}_{-m}=0)&\gauche{(3.30)}\\
&=0\mbox{ for }m=-1,0,1.
\aligner
Putting $\tilde\BY_{m,1}$ from (3.30) into the second relation of (3.28)
implies (modulo constants)
$$
\Bigl[\LR_{m,1}+\sum_{j=-\iy}^{\iy}c_j^{(m)}J^{(1)}_{j+m},\LR_{\ell,1}+
\sum_{j=-\iy}^{\iy}c_j^{(\ell)}J^{(1)}_{j+\ell}
\Bigr]=(m-\ell)(\LR_{m+\ell,1}+\Sg c_j^{(m+\ell)}J^{(1)}_{j+m+\ell}) $$
which also equals by explicit computation, using (3.12):
$$
=(m-\ell)\LR_{m+\ell,1}-\Sg c_j^{(\ell)}(j+\ell)J_{m+j+\ell}^{(1)}+\Sg
c_j^{(m)}(j+m)J^{(1)}_{m+j+\ell}.
$$
Comparing the coefficients of the $J^{(1)}$'s in two expressions on the right hand side
yields
$$
(m-\ell)c_j^{(m+\ell)}=(m+j)c_j^{(m)}-(\ell +j)c_j^{(\ell)}
\quad\mbox{provided }m+j+\ell\neq 0\leqno{(3.31)}
$$
with $c_j^{(m)}=0$ for $m=-1,0,1$, all $j\in\BZ.$

Setting $\ell =0 $ in (3.31), yields
$$
j(c_j^{(m)}-c_j^{(0)})=0\quad\mbox{and thus}\quad c_j^{(m)}=c_j^{(0)}
\quad\mbox{for}\quad j\neq 0,-m,
$$
implying
$$
c_j^{(m)}=0\quad\mbox{all }m\geq -1\quad\mbox{and}\quad j\neq 0,-m.
$$
Also, setting $j=0$ and $\ell =-1$ in (3.31) yields
$$
mc_0^{(m)}=(m+1)c_0^{(m-1)}-c_0^{(-1)},\quad\mbox{ for }m\geq 2,
$$
implying by induction, since $c_0^{(-1)}=c_0^{(1)}=0$
$$
c_0^{(m)}=0\quad\mbox{for all}\quad m\geq -1,
$$
concluding the proof.

\medbreak
{\sl Proof of Corollary 3.2.1:}  According to theorem 3.2, the vector field
$$
\BY_{m,1}+\frac{m+1}{2}\BY_{m,0}=-(ML^{m+1}+\frac{m+1}{2}L^{m})_b,
$$
acting on $\Psi$, induces on $\tau_j$ the vector field
\begin{eqnarray*}
{\cal L}_{m,1}^{j}+\frac{m+1}{2}{\cal L}_{m,0}^{j}&=&
J_{m}^{(2)}+(2j-m-1)J_{m}^{(1)}+(j^2-j)\dt_{m,0}\\
& &\quad +\frac{m+1}{2}(2J_{m}^{(1)}+2j\dt_{m,0})\\
&=&J_{m}^{(2)}+2jJ_{m}^{(1)}+j^2J_{m}^{(0)}
\end{eqnarray*}
establishing the corollary.

\medbreak

\noindent Example : symmetries at $t=0$.

By (2.4), the symmetries take on the following form
$$
\BY_{\ell,1}L=L^{\ell+1}+[(ML^{\ell+1})_s,L],
$$
with (see (1.23))
$$
M=S\vr S^{-1}+\frac{1}{2}\sum^{\infty}_1
kt_kL^{k-1}=P+\frac{1}{2}\sum^{\infty}_1 kt_kL^{k-1}.
$$
Therefore, at $t=0$,
$$
\BY_{\ell,1}L\Bigl|_{t=0}=L^{\ell+1}+[(PL^{\ell+1})_s,L],
$$
with 
$$
(PL^{\ell+1})_s=([S,\nu]\dt^{\ell}S^{-1})_s+(\nu L^{\ell})_s,
$$
upon using (3.20). In the formula above, the wave operator
$S=\ga^{-1}\tilde S$ is given by (1.9) and one finds, by a
computation similar to (3.21),
$$
(\nu L^{\ell})_s=\nu(L^{\ell})_s+[\nu,(L^{\ell})_-].
$$
If $b_i$ and
$a_i$ stand for the diagonal and
off-diagonal elements of $L$ respectively, i.e.
$L=\dt^{-1}a+b\dt^0+a\dt$, we have, in view of Lemma 3.4,
$$
\BY_{-1,1}L\Bigl|_{t=0}=I,\quad\quad\BY_{0,1}L\Bigl|_{t=0}=L
$$
and
$$
\BY_{1,1}L\Bigl|_{t=0}~:~\left\{\begin{array}{l}
\dot b_i=b_i^2+(2i+1)a_i^2-(2i-3)a^2_{i-1}\\
\dot a_i=a_i((i+1)b_{i+1}-(i-1)b_i).\end{array}\right.
$$
The subsequent symmetry vector fields can all be computed and are
non-local; for instance $\BY_{2,1}L$ involves the coefficients
$\frac{\pl}{\pl t_1}\log\tau_n=\sum^{n-1}_0 b_i$ of $\dt^{-1}$ in
$\ga^{-1}\tilde S$ (see (1.9)).

\section{Orthogonal polynomials, matrix
integrals, skew-symmetric matrices and Virasoro
constraints}

Remember from the introduction the orthogonal (orthonormal)
polynomial basis of $\HR^+=\{1,z,z^2,\ldots\}$ on the interval $[a,b],~-\iy\leq
a<b\leq\iy$,
$$
\tilde p_r(t,z)=z^r+\pp\mbox{ (monic) and
}p_r(t,z)=\frac{1}{\sqrt{h_r(t)}}\tilde p_r(t,z)\mbox{ (orthonormal),
}r\geq 0,\leqno{(4.1)}
 $$
with regard to the t-dependent inner product (via the exponential $e^{\sum
t_iz^i}$) :
$$
\langle
u,v\rangle_t=\int_a^buv\rho_tdz,~\mbox{with}~\rho_t(z)=
e^{-V_0(z)+\sum_1^{\iy}t_iz^i}=\rho_0(z)e^{\sum
t_iz^i};\leqno{(4.2)} 
$$
i.e.,
$$
\langle\tilde p_i,\tilde p_j\rangle_t =h_i\dt_{ij}\mbox{  and 
}\langle p_i,p_j\rangle_t =\dt_{ij}.
 $$
Then the semi-infinite vector (of $\la \,,\,\ra_0$-orthonormal functions)
$$
\Psi(t,z):=e^{\frac{1}{2}\Sg t_iz^i}p(t,z):=e^{\frac{1}{2}\Sg t_iz^i}(p_0(t,z),
p_1(t,z),\pp)^{\top},\leqno{(4.3)} $$
satisfies the orthogonality relations
$$
\la(\Psi(t,z))_i,(\Psi(t,z))_j\ra_0=\la
p_i(t,z),p_j(t,z)\ra_t =\dt_{ij},\leqno{(4.4)} $$
The weight is assumed to have the following property\footnote{the choice of $f_0$ 
is not unique. When $V'_0$ is rational,
then picking $f_0=$ (polynomial in the denominator) is a canonical choice.}:
$$
-\frac{\rho'_0}{\rho_0}=V'_0=\frac{\sum_0^{\iy}b_iz^i}{\sum_0^{\iy}a_iz^i}=
\frac{h_0(z)}{f_0(z)}
\quad\mbox{with}\quad\rho_0(a)f_0(a)a^k=
\rho_0(b)f_0(b)b^k=0,\quad k\geq 0. \leqno{(4.5)}
$$
Define semi-infinite matrices $ L$ and $ P$ such that
$$z p(t,z)= L(t) p(t,z),\quad\frac{\pl}{\pl
z}p(t,z)= Pp(t,z) \leqno{(4.6)}$$

The ideas of Theorem 4.1 are due to Bessis-Itzykson-Zuber [BIZ] and Witten
[W].

\proclaim Theorem 4.1. 
The semi-infinite vector $\Psi(t,z)$ and the
semi-infinite matrices $ L(t)$ (symmetric), and
$ M(t):= P(t)+\frac{1}{2}\sum_1^{\iy}kt_k L^{k-1}$, satisfy $$
z\Psi(t,z)= L(t)\Psi(t,z),\quad\mbox{and}\quad\frac{\pl}{\pl
z}\Psi(t,z)= M\Psi(t,z)\leqno{(4.7)}
$$ and
$$
\frac{\pl L}{\pl t_n}=\frac{1}{2}[( L^n)_s, L],\quad 
\frac{\pl M}{\pl t_n}=\frac{1}{2}[( L^n)_s, M],
\mbox{ and }\frac{\pl \Psi}{\pl
t_n}=\frac{1}{2}(L^n)_s,\Psi;\leqno{(4.8)} $$ the wave vector
$\Psi(t,z)$ and the $L^2$-norms $h_n(t)$ admit the
representation $$
\Psi(t,z)=e^{\frac{1}{2}\Sg
t_iz^i}\Bigl(z^n
\frac{\tau_n(t-[z^{-1}])}{\sqrt{\tau_n(t)\tau_{n+1}(t)}}\Bigr)_{n\geq
0}\mbox{ and } h_n(t)=\frac{\tau_{n+1}(t)}{\tau_n(t)}\leqno{(4.9)}
$$
with
$$
\tau_n(t)=\frac{1}{\Om_nn!}\int_{\MR_n(a,b)}e^{-Tr\,V_0(Z)+\Sg
t_iTr\,Z^i}dZ;\leqno{(4.10)} $$
the integration is taken over the space $\MR_n(a,b)$ of $n\times n$
Hermitean matrices with eigenvalues $\in [a,b].$

\medbreak
{\sl Proof:} \underline{Step 1}. Suppose $\dot\Psi={\cal B}\Psi$ and
$P(z,\frac{\pl}{\pl z})\Psi=\PR\Psi$ where $\rd$ and $\PR$ are
matrices and $P$ is a polynomial with constant coefficients. Then
$$
\dot\PR =[\rd,\PR].
$$
Indeed, this follows from differentiating $P(z,\frac{\pl}{\pl
z})\Psi=\PR\Psi$, and observing that
$$
P\rd\Psi=P\dot\Psi=\dot\PR\Psi+\PR\dot\Psi=\dot\PR\Psi+\PR\rd\Psi.
$$

\medbreak

\noindent\underline{Step 2}. The matrix $L\in\DR_{-\iy,1}$ is
symmetric because the operation of multiplication by $z$ is
symmetric with respect to $\langle\,,\,\rangle$ on $\HR^+$ and is
represented by $L$ in the basis $p_i.$ Moreover, $
P\in\DR_{-\iy,-1}$ and $[ L, P]=[ L, M]=1$. Also for $k\geq 0$,
$$\frac{\pl  \Psi}{\pl z}=\frac{\pl p}{\pl
z}e^{\frac{1}{2}\sum_1^{\iy}t_iz^i}+\frac{1}{2}\sum_1^{\iy}it_iz^{i-1}
 p e^{\frac{1}{2}\sum_1^{\iy}t_iz^i}=\left( P+\frac{1}{2}\Sigma it_i 
L^{i-1}\right)\Psi
=  M\Psi.\leqno{(4.11)}
$$
establishing (4.7).

\medbreak

\noindent\underline{Step 3}. We now prove the first statement of
(4.8). Since $\partial  p_k/{\partial} t_i$ is again a polynomial of
the same degree as $ p_{ k}$, we have:
$$ \frac{\partial  p_k}{\partial t_i}=\sum_{0\leq\ell \leq  k}
A^{(i)}_{k\ell}  p_{\ell},\,\,\,\,A^{(i)}\in\DR_b.\leqno{(4.12)}
$$
The precise nature of  $A^{(i)}$ is found as follows:
for $\ell <k,$
\begin{eqnarray*}
0&=&\frac{\partial}{\partial t_i} \int 
 p_k(z) p_{\ell} (z)\rho_t(z) dz\\
&=&\int \frac{\partial
 p_k} {\partial t_i}  p_{\ell}\rho_t(z)dz+\int   
p_k\frac{\partial  p_{\ell}}{\partial t_i} \rho_t(z)dz+\int dz (\frac{\partial}{\partial t_i} e^{-V_0+\Sg t_j
z^j})  p_k p_{\ell},\quad\mbox{using }\langle
p_i,p_j\rangle =\dt_{ij}\\ &=&\int  \sum_{j \leq k} A^{(i)}_{kj} 
p_j  p_{\ell}\rho_t(z)dz+\int \sum_j ( L^i)_{kj} 
p_j  p_{\ell}\rho_t(z) dz,~\mbox{ using (4.12)
and (4.6) }\\ &=&A^{(i)}_{k
\ell}+( L^i)_{k \ell}  
\end{eqnarray*}
and for $\ell =k$,
\begin{eqnarray*}
0=\frac{\partial}{\partial t_i} \int (
p_k(z))^2\rho_t(z)dz&=&2\int \sum_{m\leq k}A_{km}^{(i)} p_m
p_k \rho_t(z)dz +\int \sum_j(  L^i)_{kj} p_j
p_k\rho_t(z)dz\\ &=&2 A^{(i)}_{kk}+( L^i)_{kk} ,
\end{eqnarray*} 
implying
$$
A^{(i)} = -(  L^i)_- - \frac{1}{2} ( L^i)_0 = 
- \frac{1}{2}(  L^i)_b, \leqno{(4.13)}
$$
and thus
$$
\frac{\pl\Psi}{\pl t_i}=\frac{\pl}{\pl t_i} e^{\Sg
/2}p=\frac{1}{2} e^{\Sg /2} z^i p-\frac{1}{2}e^{\Sg /2}
( L^i)_b p=\frac{1}{2} ( L^i-( L^i)_b)
\Psi=\frac{1}{2} ( L^i)_s\Psi. 
$$
Now using step 1, we have immediately (4.8). So $\Psi$ satisfies the Toda equations
(1.16) and behaves asymptotically as:
$$
\Psi=e^{\frac{1}{2}\Sg
t_iz^i}\Bigl(
\frac{1}{\sqrt{h_n}}z^n(1+O(z^{-1}))\Bigr)_{n\in\BZ}
\leqno{(4.14)}$$
Therefore in view of (1.15), we must have
$$
\sqrt{h_n}=\ga_n = \sqrt{\frac{\tau_{n+1}}{\tau_n}}.
$$

\medbreak

\noindent\underline{Step 4}. The
integration (4.10) is taken with
respect to the invariant measure
$$
dZ=\prod_{1\leq i\leq n}dZ_{ii}\prod_{1\leq i< j\leq n}
d(ReZ_{ij})d(Im Z_{ij}).\leqno{(4.15)}
$$
Since the integrand only depends on the spectrum of $Z$ and since the measure
separates into an ``angular" and a ``radial" part, one first integrates out the
former, accounting for the
$\Omega_n$  and next the latter, in terms of the monic orthogonal
polynomials $\tilde p_i$:
\ALIGNER
I_n &= \Omega_n \int_{[a,b]^n} dz_1 \ldots d
z_n \prod_{1 \leq i < j \leq n} (z_i -
z_j)^2 \prod^n_{i=1}  e^{-V(z_i)},&\gauche{(4.16)}\\
&=\Omega_n \int_{[a,b]^n}  dz_1 \ldots dz_n 
(\det(\tilde p_{i-1}(z_j))_{1 \leq i
\leq j \leq n})^2 \prod^n_1 e^{-V(z_i)}\\
&=\Omega_n n ! \int_a^b dz_1\rho_t(z_1) \tilde p_0
(z_1)^2  \ldots \int_a^b
d z_n \rho_t(z_n) \tilde p_{n-1} (z_n)^2\\
&=\Omega_n n ! h_0 \ldots h_{n-1}\\
&=\Om_nn!\tau_0\frac{\tau_1}{\tau_0}\frac{\tau_2}{\tau_1}\pp
\frac{\tau_n}{\tau_{n-1}}=\Om_nn!\tau_n,\mbox{ using }\tau_0=1
\aligner
ending the proof of Theorem 4.1.

\bigbreak

Next we show that the wave vector constructed from the
orthogonal polynomial basis is a fixed point for an algebra of
symmetries, which in turn implies Virasoro-like constraints on
$\tau$. The skew-symmetry of
the matrix $Q$ below had been pointed out by E. Witten [W] in the context
of Hermite polynomials. The Virasoro constraints for the matrix
integrals with the weight $\rho_0=e^{-z^2}$ had been computed by E.
Witten [W], Gerasimov, Marshakov, Mironov, Morozov and Orlov [GMMMO]; they
relate to the deformations of Hermite polynomials. The case of deformations of
Laguerre polynomials was worked out by Haine \& Horozov [HH] and applied to
questions of highest weight representation of the Virasoro algebra.

\proclaim Theorem 4.2. Consider the semi-infinite wave vector
$\Psi(t,z)$, arising in the context of orthogonal polynomials with a weight
$\rho_t(z)$ as in (4.2) and satisfying (4.5). Then  $\Psi(t,z)$ is a fixed point for
a Lie algebra of symmetry vector fields; that is
$$
\BV_m\Psi:= -(V_m)_b\Psi =0,\quad \mbox{for}\quad m\geq -1; \leqno{(4.17)}
$$the symmetries $\BV_m$ form a (non-standard) representation of
Diff$(S^1)^+$:
$$
[\BV_m,\BV_n]=(m-n)\sum_{i\geq 0}a_i\BV_{m+n+i},\quad\quad -1\leq
n,m<\iy, \leqno{(4.18)} $$
and are defined by the semi-infinite 
matrices\footnote{in terms of the anticommutator
$(0.20)$} \footnote{set $g_0:=\frac{(f_0 \rho_0)'}{2\rho_0}=\frac{f_0'-h_0}{2}$,
with
$h_0:=-\frac{f_0\rho_0'}{\rho_0}$}
$$
V_m:=\{ Q, L^{m+1}\}= Q
L^{m+1}+\frac{m+1}{2} L^mf_0( L),~ 
\mbox{with}~
Q:= Mf_0( L)+g_0( L), \leqno{(4.19)} $$
which are skew-symmetric on the locus of $\Psi(t,z)$ above. Moreover $ Q$ is a
solution of the ``string equation"
$$
[ L, Q]=f_0( L),\leqno{(4.20)}
$$
and the $\tau$-vector satisfies the Virasoro constraints
$$
{\cal V}_m^{(n)}\tau_n=\sum_{i\geq
0}\left(
a_i(J_{i+m}^{(2)}+2n\,J_{i+m}^{(1)}+n^2\,J^{(0)}_{i+m})-b_i
(J_{i+m+1}^{(1)}+n\,J^{(0)}_{i+m+1})
\right)\tau_n=0,\leqno{(4.21)}
$$
$$\hspace{7cm} \mbox{ for}~ m=-1,0,1,\pp,~n=0,1,2,\pp ,$$
with the ${\cal V}_m^{(n)}, m\geq-1$ (for fixed $n\geq0 $) satisfying the same
Virasoro relations as (4.18), except for an additive constant. 

\bigbreak
\noindent In preparation of the proof we give some elementary lemmas.

\proclaim Lemma 4.3. Consider operators $S_1$ and $S_f$ acting on a suitable
space of functions of $z$, such that $[S_1,z]=1$ and
$S_f=\sqrt{f}S_1\sqrt{f}$; then the following
holds\footnote{$(h_1,h_2)=h_1h'_2-h_2h'_1$ denotes the Wronskian}:\newline (i)
$S_f=\{S_1,f\}$ and $[S_f,z]=f$,\newline (ii) $[\{S_1,h_1\},\{S_1,h_2\}]=\{S_1,
(h_1,h_2)\}$\newline
(iii)$\{S_1,h_1h_2\}=\{\{S_1,h_1\},h_2\}=\{\{S_1,h_2\},h_1\}$\newline (iv)
$[\{S_1,fz^{m+1}\},\{S_1,fz^{n+1}\}]=(n-m)\{S_1, f^2z^{m+n+1}\}
=(n-m)\{\{S_1,
fz^{m+n+1}\},f\}$\newline (v) the operators $\{S_1,z^{m+1}\},~m\in \BZ$ form a
representation of Diff$(S^1)$, 
$$[\{S_1,z^{m+1}\},\{S_1,z^{n+1}\}]=(n-m)\{S_1,
z^{m+n+1}\}
$$
\noindent (vi) given $f(z)=\sum_{i\geq 0}a_iz^i$, the $\{S_1,fz^{m+1}\},~m\in
\BZ$ also form a representation of Diff$(S')$:
$$[\{S_1,fz^{m+1}\},\{S_1,fz^{n+1}\}]=(n-m)\sum_{i\geq 0}a_i\{S_1,
fz^{m+n+i+1}\},$$
the map to the standard generators, $f^{-1}(z)=\sum_{i\geq -k}\bar a_iz^i$,
$$
\{S_1,fz^{m+1}\}\mapsto\{S_1,z^{m+1}\}=\{S_1,f^{-1}\cdot
fz^{m+1}\}=\sum_{i\geq -k}\bar a_i\{S_1,fz^{m+i+1}\}.
$$

{\sl Proof:} $[S_1,z]=1$ implies $[S_1,z^n]=nz^{n-1}$, since
$[S_1,.]$ is a derivation, and thus $[S_1,h]=h'$, which leads to
$$
S_f=\sqrt{f}S_1\sqrt{f}=\sqrt{f}[S_1,\sqrt{f}]+fS_1=\frac{1}{2}(2fS_1+f')=
\frac{1}{2}(2fS_1+[S_1,f])=\{S_1,f\}.$$
The second part of $(i)$,$(ii)$ and $(iii)$ follows by direct computation;
$(iv)$ is an immediate consequence of $(ii)$,
$(iii)$ and the Wronskian identity
$$(fz^{m+1},fz^{n+1})=(n-m)f^2z^{m+n+1}.$$
The Virasoro relations $(v)$ follow immediately from $(iv)$, whereas $(vi)$
follows from the argument:

$$[\{S_1,fz^{m+1}\},\{S_1,fz^{n+1}\}]=
(n-m)\{S_1,
f^2z^{m+n+1}\}$$
$$~~~~~~~~~~~~=(n-m)\{\{S_1,
fz^{m+n+1}\},f\} ,~~~\mbox{by}~~ (iii)$$
$$~~~~~~~~~~~~=(n-m)\sum_{i\geq 0}a_i\{\{S_1,
fz^{m+n+1}\},z^i\}
$$
$$~~~~~~~~~~~~~~~~~~~=(n-m)\sum_{i\geq 0}a_i\{S_1,
fz^{m+n+i+1}\},~~~\mbox{by}~~ (iii)
$$
ending the proof of lemma 4.3.

\proclaim Lemma 4.4. Consider the function space $ 
{\cal H}=\{\ldots,z^{-1},1,z,\ldots\}$ with a real inner product
$\langle u,v\rangle_{\rho} = \int^b_a uv\rho dz$, $-\iy\leq
a<b\leq\iy$  with regard to the weight
$\rho$, with $\rho(a)=\rho(b)=0$; also consider an arbitrary function
$f=\sum_{i\geq0} a_iz^i$ with
$f(a)
\rho(a) a^m=f(b)\rho(b) b^m=0$, for $m\in\BZ$. Then 
the first-order differential operator from $\cal H$ to $\cal H$
$$  S=f\frac{d}{dz}+g, $$ 
is
skew-symmetric for $\langle ~, \rangle_{\rho}$ if and only if S takes on the
form
$$ S = \sqrt{\frac{f}{\rho}} \frac{d}{dz}
\sqrt{f
\rho} = \{S_1,f\},~\mbox{where}~ S_1= \frac{1}{\sqrt\rho} \frac{d}{dz}
\sqrt{
\rho}=\frac{d}{dz}+\frac{1}{2}\frac{\rho'}{\rho}\leqno{(4.22)}
 $$
So, the operators $\{S_1,z^{m+1}\}$ and $\{S_1,fz^{m+1}\}$, for $m\in \BZ$, 
form representations of Diff$(S^1)$ in
the space of skew-symmetric operators $so({\cal H}, <,>_{\rho})$.

\medbreak
{\sl Proof:} First compute the expressions:

\begin{eqnarray*}
\langle  Su,v\rangle&=&\int_a^b f\frac{du}{dz}v\rho\,
dz+\int_a^b guv\rho dz\\
&=&u vf\rho \mid_a^b-\int_a^bu \frac{d}{dz}
(vf\rho)dz+\int_a^bguv\rho dz\\
&=&\int_a^b\rho u\Bigl[\rho^{-1}(-\frac{d}{dz}f+g)(\rho v) \Bigr]dz,\mbox{
using }f\rho(a)=f\rho(b)=0
\end{eqnarray*}
and
$$
\langle u,Sv\rangle=\int_a^b\rho u(f\frac{d}{dz}+g)
vdz.
$$
Imposing $S$ skew, i.e.,
$\langle Su,v\rangle =\langle u, S^T v\rangle  =-\langle u,Sv\rangle $,
leads to the operator identity  
$$
\rho^{-1}(-\frac{d}{dz}f+g)\rho=-(f\frac{d}{dz}+g),
$$
in turn,  leading to $g=\frac{1}{2} \rho^{-1} (f\rho)'$; thus S takes on
the form (4.22). The last part of the proof of Lemma 4.4 follows at once from the
above and Lemma 4.3
$(iv)$ and $(v).$

\bigbreak

\proclaim Lemma 4.5. Consider the above inner-product $\langle u,v\rangle$
in the space $  {\cal H}^+=\{1,z,z^2,\ldots\}$,  
for the weight $\rho$ 
having a representation of the form
$$
-\frac {\rho'}{\rho}= \frac{\sum_{i\geq 0}b_iz^i}{\sum_{i\geq 0}a_iz^i}
\equiv \frac {h}{f}~. \leqno{(4.23)}$$
Let ${\cal H}^+$ have an orthonormal basis of functions $(\varphi_k)_{k\geq
0}$;  then the operators\linebreak$\{S_1,fz^{m+1}\}$ for $m\geq-1$ are maps
from ${\cal H}^+$ to ${\cal H}^+$ and its representing matrices in that basis
(i.e., $(\langle \{S_1,fz^{m+1}\}\varphi_k,\varphi_{\ell}\rangle)_{k,\ell\geq
0}) $, for $m\geq -1$ form a closed Lie algebra $\subset
 so(0,\infty)$\footnote{$so(0,\infty)$ denotes the Lie algebra of semi-infinite
skew-symmetric matrices.}.

\noindent Remark: The operators $\{S_1,z^{m+1}\}$ do not map ${\cal
H}^+$ in ${\cal H}^+$.

\medbreak
{\sl Proof:} The operators,
$$
\{S_1,fz^{m+1}\}=\Bigl\{\{S_1,f\},z^{m+1}\Bigr\}:\HR\rg\HR,
\quad\mbox{for}\quad m\geq -1,
$$
which are skew-symmetric by Lemma 4.4, preserve the subspace $\HR^+$,
since by virtue of (4.22),
$$
\{S_1,f\}=f\frac{d}{dz}+\frac{(f\cdot\rho)'}{2\rho}=f\frac{d}{dz}
+\frac{f'-h}{2}\leqno{(4.24)}
$$
contains holomorphic series $f$ and $f'-h$, by (4.23). In a basis of functions
$(\vp_k)_{k\geq 0}$, orthonormal with respect to $\la\,,\,\ra_{\rho}$, the
corresponding matrices will also be skew-symmetric.

\medbreak
{\sl Proof of Theorem 4.2:} According to Lemma 4.5, the 
operators
$$
T_m:=
T_m^{(\rho_0,f_0)}:=\{\frac{1}{\sqrt{\rho_0}}\frac{d}{dz}\sqrt{\rho_0},
f_0z^{m+1}\}=\Bigl\{\{\frac{1}{\sqrt{\rho_0}}\frac{d}{dz}\sqrt{\rho_0},
f_0\},z^{m+1}\Bigr\},\quad m\geq -1\leqno{(4.25)} $$
map $\HR^+$ into $\HR^+$ and form an algebra with structure constants:
$$
[T_m,T_n]=(n-m)\sum_{i\geq 0}a_iT_{m+n+i},\quad m,n\geq -1.
\leqno{(4.26)}
$$
Under the map $\phi$ (Theorem 2.2), the operators $T_m$ get transformed into 
matrices $V_m=\phi(T_m)$, such that
$$
T_m\Psi(t,z)=V_m\Psi(t,z);\leqno{(4.27)}
$$
namely (see footnote {\footnotesize 15})
$$
Q:=V_{-1}=
\phi(T_{-1})=\phi\Bigl(\Bigl\{\frac{1}{\sqrt{\rho_0}}\frac{d}{dz}
\sqrt{\rho_0},f_0\Bigr\}\Bigr)= \phi \left( f_0\frac{d}{dz}+g_0 \right)=Mf_0( L)+
g_0( L)\leqno{(4.28)}
$$
and
\ALIGNER
V_m:&=\phi(T_m)&~\gauche{(4.29)}~\\
    &=\phi(\{T_{-1},z^{m+1}\})\\
&=\phi(z^{m+1}T_{-1}+\frac{1}{2}[T_{-1},z^{m+1}])\\
&=\phi(z^{m+1}T_{-1}+\frac{1}{2}[f_0\frac{d}{dz},z^{m+1}])\\
&=\phi(z^{m+1}T_{-1}+\frac{m+1}{2}f_0z^m)\\
    &= Q L^{m+1}+\frac{m+1}{2} L^mf_0( L)\\
			 &=\sum_{i\geq 0}a_i  M L^{i+m+1}+\sum_{i\geq0}\frac{(i+1)a_{i+1}-b_i}{2} 
      L^{i+m+1}+\frac{m+1}{2}\sum_{i\geq 0}a_i L^{i+m}\\
	   &=\sum_{i\geq 0}a_i( M L^{i+m+1}+\frac{i+m+1}{2} L^{i+m})-\sum_{i\geq
      0}\frac{b_i}{2} L^{i+m+1},
\aligner
where we used
$$
f_0=\sum_{i\geq 0}a_iz^i\quad\mbox{and}\quad
g_0=\frac{(f_0\rho_0)'}{2\rho_0}=\frac{1}{2} \sum_{i\geq
0}\bigl((i+1)a_{i+1}-b_i\bigr)z^i. $$
 
In addition, according to Lemma 4.5, the $z$-operators $T_m$ are skew-symmetric with
regard to $\la\,,\ra_0$ and thus form a representation of Diff$(S^1)^+$
$$
\mbox{Diff}(S^1)^+\lrg so(\HR^+,\la\,,\ra_0):=
\left\{
\begin{tabular}{l}
skew-symmetric operators\\
on $\HR^+,\la\,,\,\ra_0$
\end{tabular}
\right\},
$$
with structure constants given by (4.18). The components $e^{\sum t_iz^i} p_n(t,z)$,
$n\geq 0$ of
$\Psi(t,z)$ form an orthonormal basis of $\HR^+$,
$\la\,,\,\ra_0$, with regard to which the operators $T_m$ are represented by
semi-infinite skew-symmetric matrices; i.e., the anti-homomorphism $\phi$ restricts
to the following map
$$
\phi:so(\HR^+,\la\,,\,\ra_0)\lrg so(0,\iy)\quad\mbox{(anti-homomorphism).}
$$
Hence the matrices $V_m, ~m\geq-1$ are skew-symmetric (i.e., $(V_m)_b=0$) and thus,
using (4.29), we have
\ALIGNER
0=\BV_m\Psi&=-(V_m)_b\Psi &\gauche{(4.30)}\\
	        &=\left(\sum_{i\geq 0}a_i( M L^{i+m+1}+\frac{i+m+1}{2} L^{i+m})_b-\sum_{i\geq
      0}\frac{b_i}{2} (L^{i+m+1})_b\right)\Psi.
\aligner
In the final step, Theorem 3.2 and Corollary 3.2.1 lead to the
promised $\tau$-constraints (4.21), modulo a constant, i.e.,
$$
{\cal V}_m^{(k)}\tau_k=c_m^{(k)}\tau_k\quad\quad m\geq -1,k\geq
0.\leqno{(4.31)} $$
By (3.27), this constant is independent of $k$, i.e.,
$$
c_m^{(k)}=c_m^{(0)};
$$
upon evaluating (4.31) at $k=0$ and upon using $\tau_0=1$, we conclude
$c_m^{(0)}=0$, yielding (4.21), as claimed. Finally the map
$$
T_m\longmapsto {\cal V}_m,\quad\quad m\geq -1
$$
is an anti-homomorphism (modulo constants) by Lemma 3.1; we also have
$$
\phi:[T_{-1},z]=f_0(z)\longmapsto [ L, Q]=f_0( L),\leqno{(4.32)}
$$
which is the ``string equation", concluding the proof
of Theorem 4.2.

\medbreak

\noindent Remark 4.~:~Note that, if $f_0^{-1}=\sum_{i\geq -k}\bar a_iz^i$, the map
$$
T_m\mapsto \bar T_m=\sum_{i\geq -k}\bar a_iT_{m+i},\quad m\geq
k-1\leqno{(4.33)}
$$
sends $T_m$ into the standard representation of Diff$(S^1)$:
$$
[\bar T_m,\bar T_n]=(n-m)\bar T_{m+n},\quad m,n\geq k-1,\leqno{(4.34)}
$$
according to Lemma 4.3. 

\medbreak
\underline{Example}. In the next section we shall consider the classical
orthogonal polynomials; we consider here, for a given polynomial $q(z)$, in the
interval $[a,b]$ the weight
$$
\rho_0=(z-a)^{\al}e^{q(z)}(z-b)^{\be},\quad\mbox{with}\quad f_0=(z-a)(z-b)
\quad\mbox{and}\quad \al,\be\in\BZ,\geq 1.
$$
It implies that (see footnote {\footnotesize 15})
$$
f_0(a)\rho_0(a)=f_0(b)\rho_0(b)=0
$$
and that both $f_0$ and
$$
g_0=\frac{(f_0\rho_0)'}{2\rho_0}=z-\frac{(z-a)(z-b)}{2}(\frac{\al
}{z-a}+q'+\frac{\be}{z-b})-\frac{a+b}{2},$$ are polynomial.
Then  $ Q=Mf_0( L)+g_0( L)$ is skew-symmetric and $[
L, Q]=f_0( L).$ The Virasoro constraints (4.21) are then given by a finite
sum.

\section {Classical orthogonal polynomials}

It is interesting to revisit the classical orthogonal polynomials, from
the point of view of this analysis. As a main feature, we note that, in this case, not
only is
$ L$ (multiplication by 
$z$) symmetric and tridiagonal, but there exists another operator, a
first-order differential operator, which yields a {\it skew-symmetric and
tridiagonal matrix}. It is precisely given by the matrix
$ Q$! 

The classical orthogonal polynomials are characterized
by Rodrigues' formula,
$$
p_n = \frac{1}{K_n\rho_0}(\frac{d}{dz})^n (\rho_0 X^n),
$$
$K_n$ constant, $X(z)$ polynomial in $z$ of degree $\leq 2$, and
$\rho_0=e^{-V_0}$.Compare Rodrigues' formula for $n=1$ with the one for
$g_0$, ( see theorem 4.2, footnote {\footnotesize {15}})
$$
\frac{K_1}{2} p_1 = \frac{1}{2\rho_0}\frac{d}{dz}(\rho_0X)\quad\mbox{and}\quad
g_0= \frac{1}{2\rho_0}\frac{d}{dz}(\rho_0f_0),
$$
which leads to the natural identification
$$
g_0=\frac{K_1}{2} p_1 \mbox{ and } f_0=X
$$
and thus
$$
T_{-1}=f_0\frac{d}{dz}+g_0=X\frac{d}{dz}+\frac{K_1 p_1}{2}=X\frac{d}{dz}
+\frac{X'-XV'_0}{2}.
$$
Since both (degree $X$) $\leq 2$ and (degree $(X'-XV'_0)$) $\leq 1$, as will
appear from the table below, we have that $T_{-1}$, acting on polynomials, raises
the degree by at most $1$:
$$
T_{-1}p_k(0,z)=\sum_{i\leq k+1} Q_{ki}p_i(0,z),
$$
while, since $ Q$ is skew-symmetric,
$$
T_{-1}p_k(0,z)=(Qp(0,z))_k=- Q_{k,k-1}p_{k-1}(0,z)+
Q_{k,k+1}p_{k+1}(0,z),\leqno{(6.1)}
 $$
together with
$$
zp_k(0,z)= L_{k-1,k}p_{k-1}(0,z)+ L_{k,k}p_k(0,z)+
L_{k,k+1}p_{k+1}(0,z).\leqno{(6.2)}
 $$
This implies at the level of the flag
$$
\pp\supset W^t_{k-1}\supset W^t_k\equiv\mbox{
span}\{(\Psi)_k,(\Psi)_{k+1},\pp\}\supset W^t_{k+1}\supset\pp $$
that
$$
zW^t_k\subset W^t_{k-1}\quad\mbox{and}\quad T_{-1}W^t_k\subset W^t_{k-1}.
$$
Thus the recursion operators $z$ and $T_{-1}$ serve to characterize the
flag and so the wave vector $\Psi.$ It is interesting to speculate on considering
``$(p,q)$-cases", where, for instance,
$$
z^pW^t_k\subset W^t_{k-p}\quad\mbox{and}\quad T_{-1}W^t_k\subset W^t_{k-q}.
$$

The existence of two operators, a symmetric and a skew-symmetric one, both
represented by tridiagonal matrices, probably characterize the orthogonal
polynomials on the line. Related, it is interesting to point out a
conjecture by Karlin and Szeg\"o and a precise formulation by Al-Salam and Chihara,
were classical orthogonal polynomials are characterized by orthogonality and the
existence of a differentiation formula of the form
$$f_0(z)p'_n(z)=(\al_n z+\be_n)p_n(z)+\ga_np_{n-1}(z).$$

We now have the following
table:

\medbreak
 \small{
$$
\begin{tabular}{llll}
 &Hermite&Laguerre&Jacobi\\
$e^{-V_0(z)}dz$&$e^{-z^2}dz$&$e^{-z}z^{\alpha}dz$&
$(1-z)^{\alpha}(1+z)^{\beta}dz$\\
 & & & \\
$(a,b)$&$(-\infty,\infty)$&$(0,\infty)$&$(-1,1)$\\
 & & & \\
$T_{-1}=f_0\frac{d}{dz}+g_0$&$\frac{d}{dz}-z$&$z\frac{d}{dz}-
\frac{1}{2}(z-\alpha -1)$&$(1-z^2)\frac{d}{dz}$\\
 & & &$-\frac{1}{2}((\alpha +\beta
+2)z+(\alpha -\beta))$\\
 & & & \\
string&$[ L, Q]=1$&$[ L, Q]= L$&$[ L,
Q]=1- L^2$\\ equation& & &
\end{tabular}
$$
}

\medbreak

We now give a detailed discussion for each case:

\medbreak

(a) weight $e^{-z^2}dz.$

\noindent The corresponding (monic) orthogonal Hermite polynomials satisfy
the classic relations
$$
z\tilde p_n=\frac{n}{2}\tilde p_{n-1}+\tilde
p_{n+1}, \quad\mbox{and}\quad \frac{d}{dz}\tilde p_n=n\tilde p_{n-1}.
$$
Therefore the matrices defined by
$$
z\tilde p_n=\frac{n}{2}\tilde
p_{n-1}+\tilde p_{n+1}\hspace{1cm}(\frac{d}{dz}-z)\tilde p_n=
\frac{n}{2}\tilde p_{n-1}-\tilde p_{n+1} $$
can be turned simultaneously into symmetric and
skew-symmetric matrices $ L$ and $ Q= M- L$
respectively, by an appropriate diagonal conjugation. The
string equation reads $[ L, Q]=1$ and the matrix
integrals $\tau_n$ satisfy $$
{\cal V}_m^{(n)}\tau_n=(J_m^{(2)}+2n 
J_m^{(1)}-2J_{m+2}^{(1)}+n^2 \dt_{m,0})\tau_n=0,m=-1,0,1,\ldots
$$
upon using formula (7.12) for $a_0=1$, $b_1=2$ and all
other $a_i,b_i=0$; this captures the original case of
Bessis-Itzykson-Zuber and Witten [\,]; Witten had pointed out in his Harvard
lecture that $ M- L$ is a skew-symmetric matrix.

\medbreak

(b) weight $e^{-z}z^{\alpha}dz.$

\noindent Again the classic relations for (monic)
Laguerre polynomials,
\begin{eqnarray*}
z\tilde p_n&=&n(n+\alpha)\tilde p_{n-1}+(2n+\alpha +1)\tilde p_n+\tilde
p_{n+1}\\ z\frac{d}{dz}\tilde p_n&=&n(n+\alpha)\tilde p_{n-1}+n\tilde p_n
\end{eqnarray*}
yield symmetric and skew-symmetric matrices $ L$ and
$ Q$, after conjugation of
 \begin{eqnarray*}
z\tilde p_n&=&n(n+\alpha)\tilde p_{n-1}+(2n+\alpha +1)\tilde p_n+\tilde
p_{n+1}\\ (2z\frac{\pl}{\pl
z}-(z-\alpha
-1))\tilde p_n&=&n(n+\alpha)\tilde p_{n-1}+0.\tilde p_n-\tilde p_{n+1}.
\end{eqnarray*}
Setting $a_1=1$, $b_0=-\alpha$, $b_1=1$ and all other
$a_i=b_i=0$, yields
$$
{\cal V}_m^{(n)}\tau_n=(J_m^{(2)}+2n 
J_m^{(1)}+
\alpha J_m^{(1)}-
J_{m+1}^{(1)}+n(n+\al)\dt_{m,0})\tau_n=0,m=0,1,2,\ldots ,$$
and the string equation $[ L, Q]= L$; this case was investigated by
 Haine and Horozov [HH].

 \medbreak

(c) weight $(1-z)^{\alpha}(1+z)^{\beta}dz.$

\noindent The matrices $L$ and $Q$ will be defined by the operators acting
on (monic) Jacobi polynomials 
 \begin{eqnarray*}
z\tilde p_n&=&A_{n-1}\tilde p_{n-1}+B_n\tilde p_n+\tilde p_{n+1}\\
-(\frac{1}{n+1}+\frac{\alpha
+\beta}{2})^{-1}((1-z^2)\frac{d}{dz}+\frac{(f_0\rho_0)'}{2\rho_0})\tilde p_n
&=&-A_{n-1}\tilde p_{n-1}+\tilde p_{n+1}
\end{eqnarray*}
with
\begin{eqnarray*}
A_{n-1}&=&\frac{4n(n+\alpha
+\beta)(n+\alpha)(n+\beta)}{(2n+\alpha
+\beta)^2(2n+\alpha +\beta +1)(2n+\alpha +\beta -1)}\\
B_n&=&-\frac{\alpha^2-\beta^2}{(2n+\alpha
+\beta)(2n+\alpha +\beta +2)}.
\end{eqnarray*}
Setting
$$
a_0=1,a_1=0,a_2=-1,b_0=\alpha -\beta,b_1=\alpha+\beta
$$
and all other $a_i=b_j=0$ leads to the constraints
$$
(J_m^{(2)}-J_{m-2}^{(2)}-2n
J_{m-2}^{(1)}+2n
J_m^{(1)}+(\alpha
-\beta) J_{m-1}^{(1)}+(\alpha
+\beta) J_m^{(1)}-n^2\dt_{m,2}+n(\al-\be)\dt_{m,1})\tau_n=0,~m=1,2,3,\ldots $$
and the string equation $[ L, Q]=1- L^2$. Gegenbauer
($\alpha=\beta=\lambda-1/2$) and Legendre  ($\alpha=\beta=0$) polynomials are
limiting cases of Jacobi polynomials. 

\section {Appendix: Virasoro constraints via the integrals}

The Virasoro constraints for the integrals can be shown in a direct way:

\proclaim Lemma A.1. Given $f_0(z)=\sum_{j\geq 0}a_jz^j$, the
following holds for $k\geq -1$:
$$
e^{\sum_1^{\iy}t_iZ^i}\frac{\pl}{\pl\vr}e^{\vr
f_0(Z)Z^{k+1}\frac{\pl}{\pl Z}}dZ\Bigl|_{\vr=0}=dZ\sum_{r\geq 0}a_r
\Bigl(\sum_{i+j=r+k}
\frac{\pl^2}{\pl t_i\pl
t_j}+2n\frac{\pl}{\pl
t_{r+k}}+n^2 \delta_{r+k}\Bigr)e^{\sum_1^{\iy}t_iZ^i},\leqno{(A.1)} $$
where $\pl/\pl t_j=0$ for $j\leq 0$.

\medbreak
{\sl Proof:} We break up the proof of this Lemma in elementary
steps, involving the diagonal part of $dZ$, i.e.,
$$
dZ=dz_1\pp dz_n\Delta(z)^2{\bf \times}\mbox{ angular part.}
$$
At first, we compute for $k\geq 0$:
$$
\frac{2}{\Delta(z)}\frac{\pl}{\pl\vr}\Bigl(e^{a\vr\sum_{i=1}^nz_i^{k+1}\frac{\pl}{\pl
z_i}}\Bigr)\Delta(z)\leqno{(A.2)}
$$
\begin{eqnarray*}
&=&2\frac{\pl}{\pl\vr}\log\Delta(z_1+a\vr z_1^{k+1},\pp,z_n+a\vr
z_n^{k+1})\Bigl|_{\vr=0}\\
&=&2a\Biggl(\renewcommand{\arraystretch}{0.5}
\begin{array}[t]{c}
\sum\\
{\scriptstyle 1\leq\al <\be\leq n}\\
{\scriptstyle i+j=k}\\
{\scriptstyle i,j> 0}
\end{array}
\renewcommand{\arraystretch}{1}
z_{\al}^iz_{\be}^j+(n-1)\sum_{1\leq\al\leq n}z_{\al}^k\Biggr)-an(n-1)\delta_{k,o}
\end{eqnarray*}
and for $k\geq -1$
$$
\frac{\frac{\pl}{\pl\vr}e^{a\vr\sum_{i=1}^nz_i^{k+1}\frac{\pl}{\pl
z_i}}dz_1\pp
dz_n\Bigl|_{\vr=0}}{dz_1\pp dz_n}\leqno{(A.3)}
$$
\begin{eqnarray*}
&=&a\frac{\pl}{\pl\vr}\prod^n_{\al=1}(1+\vr(k+1)
z_{\al}^k)\Bigl|_{\vr=0}\\
&=&a(k+1)\sum_{1\leq\al\leq n}z_{\al}^k.
\end{eqnarray*}
Note that both expressions (A.2) and (A.3) vanish for k=-1. Also, we have for $k\geq
1$,
$$
\frac{\renewcommand{\arraystretch}{0.5}
\begin{array}[t]{c}
\sum\\
{\scriptstyle i+j=k}\\
{\scriptstyle i,j> 0}
\end{array}
\renewcommand{\arraystretch}{1}\frac{\pl^2}{\pl t_i\pl t_j}
e^{\renewcommand{\arraystretch}{0.5}
\begin{array}[t]{c}
\sum\\
{\scriptstyle 1\leq i \leq \infty}\\
{\scriptstyle 1\leq \al \leq n}
\end{array}t_i z^i_{\al}}
}
{~e^{\renewcommand{\arraystretch}{0.5}
\begin{array}[t]{c}
\sum\\
{\scriptstyle 1\leq i \leq \infty}\\
{\scriptstyle 1\leq \al \leq n}
\end{array}t_i z^i_{\al}}
}
=2\renewcommand{\arraystretch}{0.5}
\begin{array}[t]{c}
\sum\\
{\scriptstyle 1\leq\al<\be\leq n}\\
{\scriptstyle i+j=k}\\
{\scriptstyle i,j> 0}
\end{array}
\renewcommand{\arraystretch}{1}
z^i_{\al}z^j_{\be}+(k-1)\sum_{1\leq\al\leq
n}z_{\al}^k\leqno{(A.4)} $$
and
$$
(2n\frac{\pl}{\pl
t_k}\log)~e^{\renewcommand{\arraystretch}{0.5}
\begin{array}[t]{c}
\sum\\
{\scriptstyle 1\leq i \leq \infty}\\
{\scriptstyle 1\leq \al \leq n}
\end{array}t_i z^i_{\al}}=2n\sum_{1\leq\al\leq n}z_{\al}^k.\leqno{(A.5)} $$
Summing up (A.2) and (A.3) yields for $k\geq 0$:
$$
\frac{\frac{\pl}{\pl\vr}e^{a\vr\sum_{i=1}^nz_i^{k+1}\frac{\pl}{\pl
z_i}}\Delta^2(z)dz_1\pp
dz_n\Bigl|_{\vr=0}}{\Delta^2(z)dz_1\pp
dz_n}=a\left(2\renewcommand{\arraystretch}{0.5} \begin{array}[t]{c}
\sum\\
{\scriptstyle 1\leq\al<\be\leq n}\\
{\scriptstyle i+j=k}\\
{\scriptstyle i,j> 0}
\end{array}
\renewcommand{\arraystretch}{1}
z^i_{\al}z^j_{\be}+(2n+k-1)\sum_{1\leq\al\leq
n}z_{\al}^k-n(n-1)\delta_k\right);
$$
this expression vanishes for $k=-1$. This expression equals  the sum of (A.4) and
(A.5); thus
$$
\frac{\frac{\pl}{\pl\vr}e^{a\vr\sum_{i=1}^nz_i^{k+1}\frac{\pl}{\pl
z_i}}\Delta^2(z)dz_1\pp
dz_n\Bigl|_{\vr=0}}{\Delta^2(z)dz_1\pp
dz_n}=\frac{a\left(\renewcommand{\arraystretch}{0.5}
\begin{array}[t]{c}
\sum\\
{\scriptstyle i+j=k}\\
{\scriptstyle i,j> 0}
\end{array}
\renewcommand{\arraystretch}{1}
\frac{\pl^2}{\pl t_i\pl t_j}+2n\frac{\pl}{\pl
t_k}+\delta_k n^2 \right)e^{\renewcommand{\arraystretch}{0.5}
\begin{array}[t]{c}
\sum\\
{\scriptstyle 1\leq i \leq \infty}\\
{\scriptstyle 1\leq \al \leq n}
\end{array}t_i z^i_{\al}}}
{e^{\renewcommand{\arraystretch}{0.5}
\begin{array}[t]{c}
\sum\\
{\scriptstyle 1\leq i \leq \infty}\\
{\scriptstyle 1\leq \al \leq n}
\end{array}t_i z^i_{\al}}}
$$
establishing Lemma A.1.

\proclaim Theorem A.2. Let the weight $\rho_0=e^{-V_0}$ have a
representation (not necessarily unique) of the form
$$
V'_0=\frac{\sum_{i\geq 0}b_iz^i}{\sum_{i\geq 0}a_iz^i}
\equiv\frac{h_0}{f_0}.
$$
Then the matrix integral $\tau_n=\frac{I_n}{\Om_nn!}$ satisfies the
KP equation, and  Virasoro-like constraints:
$$\sum_{i\geq
0}\left(
a_i(J_{i+m}^{(2)}+2n\,J_{i+m}^{(1)}+n^2\,J^{(0)}_{i+m})-b_i
(J_{i+m+1}^{(1)}+n\,J^{(0)}_{i+m+1})
\right)\tau_n=0\quad\mbox{with }m\geq -1.$$

\medbreak
{\sl Proof:} Shifting the integration variable $Z$ by means of
$$ Z\mapsto Z+\vr f_0(Z)Z^{m+1},\quad m\geq -1
$$
and using the notation
$$
\Phi(Z)=e^{-Tr\,V_0(Z)+\Sg t_iTr\,Z^i},
$$
we compute, since the integral remains unchanged, that
\begin{eqnarray*}
0&=&\frac{\pl}{\pl\vr}\int_{\MR_n}e^{\vr
f_0(Z)Z^{m+1}\frac{\pl}{\pl Z}}\Phi(Z)dZ\Bigl|_{\vr=0}\\
&=&\frac{\pl}{\pl\vr}\int_{\MR_n}e^{-Tr\,V_0(Z+\vr
f_0(Z)Z^{m+1})}e^{\sum^{\iy}_{\ell=1}t_{\ell}Tr(Z+\vr
f_0(Z)Z^{m+1})^{\ell}}e^{\vr
f_0(Z)Z^{m+1}\frac{\pl}{\pl Z}}(dZ)\Bigl|_{\vr=0}\\
&=&\frac{\pl}{\pl\vr}\int_{\MR_n}e^{\vr(-Tr\,V'_0(Z)
f_0(Z)Z^{m+1}+\sum_{\ell=1}^{\iy}\ell t_{\ell}Tr\,f_0(Z)Z^{m+\ell})
+O(\vr^2)}\Phi(Z)e^{\vr f_0(Z)Z^{m+1}\frac{\pl}{\pl
Z}}dZ\Bigl|_{\vr=0}\\
&=&\int_{\MR_n}\Biggl(-Tr\,h_0(Z)Z^{m+1}+\sum^{\iy}_{\ell=1}\ell
t_{\ell}Tr\,f_0(Z)Z^{m+\ell}\\
& &\hspace{2cm}+\sum_{i\geq 0}a_i
\Bigl(\sum_{\al+\be=i+m}\frac{\pl^2}{\pl t_{\al}\pl
t_{\be}}+2n\frac{\pl}{\pl t_{i+m}}+n^2\dt_{i+m}\Bigr) \Biggr)\Phi(Z)dZ\\
& &\hspace{5cm}\mbox{using Lemma A.1.}\\
&=&\int_{\MR_n}\Biggl(-\sum_{i\geq 0}b_iTr\,Z^{i+m+1}\\
& &\hspace{2cm}+\sum_{i\geq
0}a_i\Bigl(\sum_{\ell=1}^{\iy}\ell t_{\ell}Tr\,Z^{i+m+\ell}+
\sum_{\al+\be=i+m}\frac{\pl^2}{\pl t_{\al}\pl
t_{\be}}+2n\frac{\pl}{\pl t_{i+m}}+n^2\dt_{i+m}\Bigr)\Biggr)\Phi(Z)dZ\\
&=&\int_{\MR_n}\Biggl(-\sum_{i\geq 0,i>-m-1}b_i\frac{\pl}{\pl
t_{i+m+1}}-nb_0\dt_{m+1}\\ & &\hspace{2cm}+\sum_{i\geq
0}a_i\Bigl(\sum_{\ell=1}^{\iy}\ell t_{\ell}\frac{\pl}{\pl
t_{i+m+\ell}}+\sum_{\al+\be=i+m}
\frac{\pl^2}{\pl t_{\al}\pl t_{\be}}+2n\frac{\pl}{\pl
t_{i+m}}+n^2\dt_{i+m}\Bigr)\Biggr)\Phi(Z)dZ\\
&=&\sum_{i\geq
0}\left(
a_i(J_{i+m}^{(2)}+2n\,J_{i+m}^{(1)}+n^2\,J^{(0)}_{i+m})-b_i
(J_{i+m+1}^{(1)}+n\,J^{(0)}_{i+m+1})
\right)\tau_n,
\end{eqnarray*}
ending the proof of Theorem A.2.

\end{document}